\documentclass[10pt, final, journal, letterpaper, twocolumn]{IEEEtran}
\makeatletter
\def\ps@headings{%
\def\@oddhead{\mbox{}\scriptsize\rightmark \hfil \thepage}%
\def\@evenhead{\scriptsize\thepage \hfil \leftmark\mbox{}}%
\def\@oddfoot{}%
\def\@evenfoot{}}
\makeatother 
\IEEEoverridecommandlockouts

\usepackage{graphicx}
\usepackage{amsmath,amssymb}

\usepackage{algorithm}
\usepackage{algorithmic}
\usepackage{amsmath}
\usepackage{graphics}
\usepackage{epsfig}
\usepackage{color}
\usepackage{subfigure}
\usepackage{multicol}

\newtheorem{theorem}{Theorem}
\newtheorem{corollary}{Corollary}

\usepackage[numbers,sort&compress]{natbib}
\usepackage{flafter}

\begin{document}
\title{Secrecy Transmission in Large-Scale UAV-Enabled Wireless Networks}
\author{\IEEEauthorblockN{Jianping~Yao,~\IEEEmembership{Member,~IEEE},
and
Jie~Xu,~\IEEEmembership{Member,~IEEE}
}
\thanks
{
J. Yao, and J. Xu are with School of Information Engineering, Guangdong University of Technology, Guangzhou, P. R. China (Emails: yaojp@gdut.edu.cn, jiexu@gdut.edu.cn).
\emph{(Corresponding author: J. Xu.)}
}
}

\maketitle

\begin{abstract}
This paper considers the secrecy transmission in a large-scale unmanned aerial vehicle (UAV)-enabled wireless network, in which a set of UAVs in the sky transmit confidential information to their respective legitimate receivers on the ground, in the presence of another set of randomly distributed suspicious ground eavesdroppers.
We assume that the horizontal locations of legitimate receivers and eavesdroppers are distributed as two independent homogeneous Possion point processes (PPPs), and each of the UAVs is positioned exactly above its corresponding legitimate receiver for efficient secrecy communication. Furthermore, we consider an elevation-angle-dependent line-of-sight (LoS)/non-LoS (NLoS) path-loss model for air-to-ground (A2G) wireless channels and employ the wiretap code for secrecy transmission. Under such setups, we first characterize the secrecy communication performance (in terms of the connection probability, secrecy outage probability, and secrecy transmission capacity) in mathematically tractable forms, and accordingly optimize the system configurations (i.e., the wiretap code rates and UAV positioning altitude) to maximize the secrecy transmission capacity, subject to a maximum secrecy outage probability constraint.
Next, we propose to use the secrecy guard zone technique for further secrecy protection, and analyze the correspondingly achieved secrecy communication performance.
Finally, we present numerical results to validate the theoretical analysis.
It is shown that the employment of secrecy guard zone significantly improves the secrecy transmission capacity of this network, and the desirable guard zone radius generally decreases monotonically as the UAVs' and/or the eavesdroppers' densities increase.
\end{abstract}

\begin{IEEEkeywords}
UAV communications, physical layer security, homogeneous Poisson point process (PPP), secrecy transmission capacity, secrecy guard zone.
\end{IEEEkeywords}

\newcommand{\mv}[1]{\mbox{\boldmath{$ #1 $}}}

\IEEEpeerreviewmaketitle
\section{Introduction}
Unmanned aerial vehicles (UAVs) or drones are envisioned to have a wide range of commercial and civilian applications, such as cargo delivery, rescue and search, filming, surveillance, and aerial communication/charging platforms (see, e.g., \cite{ZengWireless2016,BerghLTE2016,YalinizThe2016,XiaoEnabling2016,MenouarUAV2017,MozaffariWireless2017} and the references therein). To enable these new applications, how to integrate UAVs into wireless communications networks is becoming an emerging topic for the wireless communications society. On one hand, UAVs can be connected with terrestrial wireless communications networks (e.g., cellular networks) as users for not only ultra-reliable and low-latency non-payload command and control but also high-speed payload data transmission \cite{ZengCellular2019,ZhangCellular2019}. On the other hand, UAVs can be employed as aerial communication platforms (such as base stations (BSs) or relays) to serve subscribers on the ground in, e.g., emergency situations, cell edges, and temporary hotspots \cite{ZengWireless2016,BerghLTE2016,Wu2018Capacity}.

The design of UAV-enabled wireless communications introduces various technical challenges due to the following differences from conventional terrestrial wireless communications. First, air-to-ground (A2G) wireless channels usually possess strong line-of-sight (LoS) components, especially when the elevation angle becomes large. Such strong LoS links are beneficial in offering better channel conditions for efficient UAV communications; however, they also result in strong co-channel interference that is harmful and thus should be properly managed. Next, UAVs have fully controllable mobility in the three-dimensional (3D) space. Although this introduces new challenges in mobility management for wireless networks, such a feature can also be exploited with proper UAV positioning/trajectory design for performance optimization (see, e.g., \cite{ZengWireless2016,ZhangCellular2019,Wu2018Capacity,HouraniOptimal2014,ChenOptimal2017,XuUAV2018,LiPlacement2018,Wang2018Coverage,Peng2018UAV} and the references therein).
Furthermore, security is another key issue faced in UAV-enabled wireless networks. Compared with conventional terrestrial communications, the A2G communication links are more vulnerable to be intercepted. Due to the existence of strong LoS components over A2G links, the UAVs' transmitted confidential data is more likely to be overheard by suspicious eavesdroppers over a large area on the ground. In this paper, we focus on the secrecy transmission in UAV-enabled wireless networks.

Over recent years, physical layer security has emerged as a viable new solution to secure wireless communications against suspicious eavesdropping attacks, which is a good complementary to conventional cryptography-based anti-eavesdropping techniques (see, e.g., \cite{YaoSecure2016,Liu2017Enhancing,YanSecret2018,WangSurvey2018,ZhengPhysical2018} and the references therein). In the physical layer security design, the secrecy rate is a widely-adopted performance metric, which is defined as the achievable rate of the confidential information transmitted over a wireless channel, provided that the eavesdroppers cannot decode any such information. It is well-established in \cite{Wyner1975,Csiszar1978} that for a Gaussian wiretap channel, the maximum secrecy rate or secrecy capacity corresponds to the difference of the legitimate channel's Shannon capacity and the eavesdropping channel's.
However, achieving such perfect secrecy requires the transmitter to perfectly know the instantaneous channel state information (CSI) with both legitimate receivers and illegitimate/suspicious eavesdroppers, which is quite challenging and even infeasible in practical wireless systems.
To tackle this issue and considering that the transmitter only knows the stochastic CSI (distributions of the channels), the secrecy outage rate/capacity becomes a favorable design metric, which is defined as the achievable communication rate of confidential information, provided that the secrecy outage probability (i.e., the probability that such information is eavesdropped) is less than a certain threshold (see, e.g., \cite{Zhou2011,zhou2013rethinking,YaoSecure2018} and the references therein).

Motivated by the great success in terrestrial communications, using physical-layer security to protect the confidentiality of UAV-enabled wireless communications has attracted growing research interest recently.
In the literature, there are generally two lines of research that focused on resource allocation for maximizing the secrecy communication performance at the {\it link} level \cite{CuiRobust2018,LiRobust2018,WangImproving2017,ZhouSecrecy2017,LeeUAV2018,ZhangSecuring2018,ZhongSecure2018,ZhaoCaching2018,CaiDual2018}, and network performance analysis and design by using stochastic geometry at the {\it network} level \cite{ZhuSecrecy2018}.
At the link level, prior works \cite{CuiRobust2018,LiRobust2018,WangImproving2017,ZhouSecrecy2017,LeeUAV2018,ZhangSecuring2018,ZhongSecure2018,ZhaoCaching2018,CaiDual2018} exploited the UAVs' controllable mobility via trajectory design, jointly with wireless resource allocation, to maximize the UAV's secrecy rate or secrecy outage rate. At the network level, to our best knowledge, \cite{ZhuSecrecy2018} is the only work that investigated the average secrecy rate of large-scale UAV-enabled millimeter wave (mmWave) networks, in which UAVs are distributed as a Mat\'{e}rn hardcore point process and a part of them are employed for cooperative jamming.
Despite such research progress, the fundamental secrecy communication performance (e.g., with secrecy outage consideration) of large-scale UAV-enabled wireless networks still remains not well addressed. This thus motivates our investigation in this work.

From the technical or stochastic geometry perspective, the analysis and design of large-scale secrecy UAV-enabled wireless networks are very challenging due to the following new considerations that are different from conventional terrestrial wireless communications (e.g., conventional cellular networks).
First, as UAVs can adjust their 3D locations for performance optimization, the point process representing the UAVs' horizontal locations is generally correlated with that of the ground nodes' (GNs') locations. Due to such correlation, the network analysis becomes a very difficult task.
Furthermore, due to the additional degree of freedom in the UAV attitude, the locations of UAVs do not follow a 2D homogeneous point process, thus making the conventional analysis approaches not applicable.
Next, A2G wireless channels generally depend on the elevation angles between UAVs and GNs. If the elevation angle is large, then the A2G channel is more likely to have strong LoS components; while if it is small, then the A2G channel is more likely to be non-LoS (NLoS) due to the obstacles between them (see, e.g., \cite{HouraniOptimal2014,KhawajaSurvey2019}). Such elevation-angle-dependent LoS/NLoS wireless channels make the network analysis even more difficult.

To address the above issues, in this paper we consider a large-scale 3D UAV-enabled wireless network, in which a set of quasi-stationary UAVs in the sky transmit confidential information to their respective legitimate ground receivers, in the presence of another set of randomly distributed suspicious ground eavesdroppers.
We suppose that all UAVs are positioned at the same altitude, as commonly assumed in the UAV communication literature \cite{LeeUAV2018,ZhangJoint2018,ZhuSecrecy2018,ChetlurDownlink2017}. Furthermore, we model the horizontal locations of legitimate receivers and suspicious eavesdroppers as two independent homogeneous Poisson point processes (PPPs) \cite{HouMultiple2018,ZhangSpectrum2017,AzariUltra2018,WuCooperative2018,GapeyenkoFlexible2018,hou2019exploiting}.
As UAVs can adjust its positioning locations for secrecy communication performance optimization, we suppose that each of the UAV transmitters is positioned exactly above its corresponding ground legitimate receiver, which can help improve the secrecy communication rate in our considered scenario.
Accordingly, the UAVs' horizontal locations are modeled as the homogeneous PPP that is identical to the legitimate receivers.
In addition, we consider an elevation-angle-dependent LoS/NLoS channel model for A2G wireless channels, such that the A2G channel is LoS when the elevation angle is larger than a given threshold, while it is NLoS otherwise. This model is consistent with the practically measured dual-slope path-loss model for A2G channels \cite[(4)]{KhawajaSurvey2019}.
Furthermore, the LoS consideration at large elevation angle also follows the A2G channel models specified by 3GPP \cite{3GPPTR36}, in which the LoS probability becomes one as the elevation angle becomes larger than a given threshold (see \cite[Table B-1]{3GPPTR36} for details).

Under such setups, the main results of this paper are summarized as follows.

\begin{itemize}
	\item First, we analytically derive the exact expressions for the connection probability, secrecy outage probability, and secrecy transmission capacity of this network. However, these expressions are mathematically too complicated to draw insights. To tackle this issue, we propose to approximate each LoS A2G channel as a Rayleigh-fading channel with the same path loss. Accordingly, we obtain the connection probability, secrecy outage probability, and secrecy transmission capacity in mathematically tractable forms. Building upon such analysis, we optimize the system configurations (i.e., the wiretap code rates and UAV positioning altitude) to maximize the secrecy transmission capacity subject to a maximum secrecy outage probability constraint.
    \item Next, we propose to use the secrecy guard zone technique for further secrecy protection, such that each UAV only communicates when there are no eavesdroppers within this zone of a certain radius. In this case, we obtain mathematically-tractable analytic expressions of the connection probability, secrecy outage probability, and secrecy transmission capacity, by similarly using the above approximations for LoS channels. Furthermore, we optimize the wiretap code rates, UAV positioning altitude, and guard zone radius, in order to maximize the secrecy transmission capacity.
    \item Finally, we present numerical results to validate the above theoretical analysis. It is shown that the employment of secrecy guard zone significantly improves the secrecy transmission capacity of this network, and the desirable guard zone radius generally decreases monotonically as the UAVs' and/or the eavesdroppers' densities increase. It is also shown that the UAVs should be positioned at the lowest altitude for maximizing the secrecy transmission capacity.
\end{itemize}

The remainder of this paper is organized as follows. Section \uppercase\expandafter{\romannumeral2} presents the system model of our considered 3D large-scale UAV-enabled wireless network.
Section \uppercase\expandafter{\romannumeral3} analyzes the secrecy communication performance of this network.
Section \uppercase\expandafter{\romannumeral4} optimizes the system configurations to maximize the secrecy transmission capacity.
Section \uppercase\expandafter{\romannumeral5} proposes to use the secrecy guard zone technique for further secrecy protection.
Section \uppercase\expandafter{\romannumeral6} presents numerical results.
Finally, Section \uppercase\expandafter{\romannumeral7} concludes this paper.

\begin{figure}[!h]
\centering
  \includegraphics[width=6.8cm]{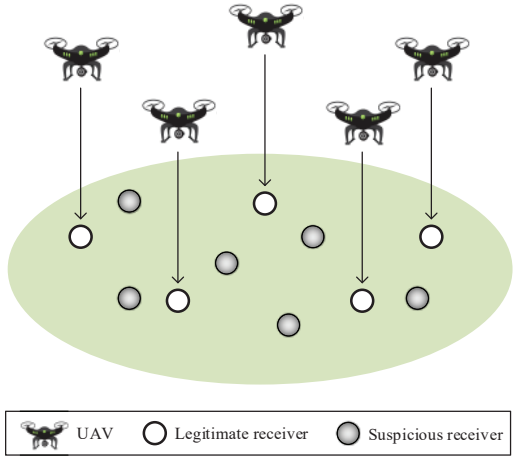}\\
  \caption{Illustration of the 3D large-scale UAV-enabled wireless network.}\label{fig:system_model}
\end{figure}

{\bfseries Notations:} $\mathbb{R}$ denotes the set of all real numbers; $\| \cdot \|$ denotes the Euclidean norm; $\mathcal{P}\left(\cdot\right)$ denotes the probability operator; $\mathbb{E}\left(\cdot\right)$ stands for the statistical expectation of a random variable;
$\mathbb{W}_0\left(\cdot\right)$ denotes the principal branch of Lambert W function; $\arcsin \left( \cdot \right)$ is the arc sine function; $\cot \left( \cdot \right)$ is the cotangent function; $\arctan \left( \cdot \right)$ is the arc tangent function.

\section{System Model}
In this work, we consider a 3D large-scale UAV-enabled wireless network as shown in Fig. \ref{fig:system_model}, in which a large number of randomly distributed UAVs in the sky transmit confidential information to their respective legitimate receivers on the ground, in the presence of randomly distributed suspicious ground eavesdroppers.{\footnote{In this paper, we consider the eavesdroppers are randomly distributed on the ground, as their locations may be uncertain in practice. For example, the eavesdroppers can be users within the same network, which do not have the right to access their communicated messages; the eavesdroppers can also be malicious nodes outside the network. In both cases, the UAVs may not know the exact locations of these eavesdroppers but only know their distribution information. As a result, it is practically relevant to assume that the eavesdroppers are randomly distributed on the ground. Actually, this is also a commonly adopted assumption in the physical-layer security literature for large-scale network performance analysis \cite{YaoSecure2016,Liu2017Enhancing,ZhengPhysical2018,Zhou2011,zhou2013rethinking,YaoSecure2018,ZhuSecrecy2018}.}
We model the horizontal locations of ground legitimate receivers and eavesdroppers as two independent homogeneous PPPs with density ${\lambda_u}$ and ${\lambda _e}$, which are denoted by ${\Phi_{u}}$ and ${\Phi _{e}}$, respectively.{\footnote{Modeling the distributions of legitimate receivers and eavesdroppers in terms of PPP has been proved very helpful for analyzing large-scale wireless communication networks, which not only naturally captures the randomness of the eavesdroppers' locations, but also is very tractable because of its powerful mathematical tools. It provides a probabilistic way of estimating spatial averages that generally capture the key dependencies of the network performance characteristics (connectivity, capacity, etc.) as functions of a relatively small number of parameters. This helps provide insights and facilitate the system design \cite{Baccelli2010}.}}
It is assumed that each UAV is positioned exactly above the corresponding legitimate receiver for improving the secrecy communication performance.{\footnote{It is shown in \cite{ZhangSecuring2018,LiRobust2018,ZhongSecure2018,CaiDual2018} that when the eavesdroppers' locations are given and known {\it a-priori}, the UAV should be positioned at the opposite direction of these eavesdroppers to maximize the secrecy rate. In our considered scenario with eavesdroppers randomly distributed over all directions, it is generally desirable to position the UAV exactly above the legitimate receiver to maximize the secrecy communication performance.}} Furthermore, for ease of exposition and as commonly adopted in the UAV communication literature \cite{LeeUAV2018,ZhangJoint2018,ZhuSecrecy2018,ChetlurDownlink2017}, it is assumed that all UAVs are positioned at the same altitude $H >0$, which is a design variable to be optimized later. Let $H_\textrm{max}$ and $H_\textrm{min}$ denote the UAVs' maximum and minimum altitudes, respectively. We thus have $H_\textrm{min} \le H \le H_\textrm{max}$.
Furthermore, it is assumed that each UAV-and-legitimate-receiver pair does not know the exact locations and CSI of other UAV-and-legitimate-receiver pairs and those of eavesdroppers, but only knows the corresponding statistical information (e.g., the eavesdroppers' density\footnote{If the eavesdroppers are active in transmissions, then UAVs can monitor their transmission to estimate the corresponding density. On the other hand, if the eavesdroppers are passive, UAVs can still detect the eavesdroppers under different scenarios. In the first scenario, the eavesdroppers can belong to the same network as the legitimate nodes, but do not have the right to access their communicated messages. In this scenario, UAVs can obtain the knowledge of $\lambda_e$ from the network controller directly. In the other scenario, the eavesdroppers can be malicious nodes outside the network. In this scenario, UAVs can detect the passive eavesdropping based on the local oscillator power leaked from the eavesdropper's RF front end \cite{Mukherjee2012}, and accordingly estimate the value of $\lambda_e$.}).

\begin{figure}[!h]
\centering
  \includegraphics[width=6.8cm]{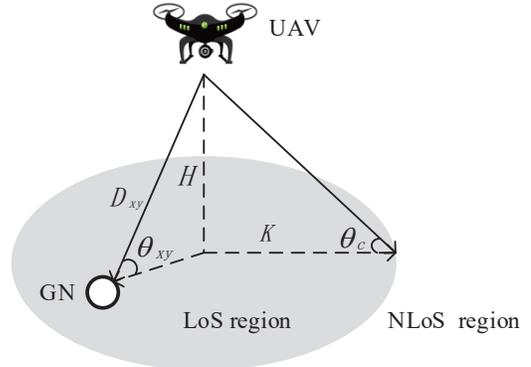}\\
  \caption{Illustration of the considered elevation-angle-dependent LoS/NLoS path-loss model.}\label{fig:channel_model}
\end{figure}

As for the A2G wireless channel, we consider an elevation-angle-dependent LoS/NLoS path-loss model with a angle threshold $\theta_c$ as shown in Fig. \ref{fig:channel_model}, which is explained in detail as follows. Consider any one particular communication link from UAV transmitter $x$ at position $\mv w_x \in \mathbb{R}^{3\times 1}$ to GN receiver $y$ (i.e., legitimate receiver or eavesdropper) at position $\mv w_y \in \mathbb{R}^{3\times 1}$, in which we use ${D_{{xy}}} = \|\mv w_x - \mv w_y\|$ to denote their distance, ${\theta_{xy}} = \frac{180}{\pi}\arcsin\left(\frac{H}{{D_{{xy}}}}\right)$ denotes the corresponding elevation angle.
In this case, if the elevation angle $\theta_{xy}$ is larger than the threshold ${\theta_{c}}$ (or, equivalently, the GN is located within a disk with radius $K = {H}\cot\left(\theta_c\right)$), then the A2G link is assumed to be LoS dominated, for which the channel power gain is dominated by the path loss, denoted by ${\eta_{_L}}{D_{xy}^{-\alpha_{_L}}}$. Here, ${\eta_{_L}}$ denotes the LoS channel power gain at the reference distance of one meter, and ${\alpha_{_L}}$ denotes the path-loss exponent in the LoS case with ${\alpha_{_L}} = 2$.
Otherwise, if the elevation angle $\theta_{xy}$ is no larger than the threshold $\theta_c$ (or, equivalently, the GN is outside the disk with radius $K$), then the A2G channel only has NLoS links, for which the channel power gain consists of the pathloss and a Rayleigh fading coefficient, denoted by ${\eta_{_N}}{S_{N}}{D_{xy}^{-\alpha_{_N}}}$, where ${\eta_{_N}}$ denotes the NLoS channel power gain at the reference distance of one meter, ${\alpha_{_N}}$ denotes the path-loss exponent in the NLoS case with ${\alpha_{_N}}= 4$ , and $S_{N}$ is an exponentially distributed random variable with unit mean. By combining the two cases and denoting $P_t>0$ as the transmit power by UAV $x$, the received signal power at GN $y$ is given by
\begin{equation}
 {P_r} = {{{\eta_{_{xy}}}{P_t}{S_{xy}}}}{{{D_{xy}^{-\alpha_{xy}}}}},
\label{received_power}
\end{equation}
where $\eta_{_{xy}}$ denotes the channel power gain at the reference distance of one meter from UAV $x$ to GN $y$ with $\eta{_{xy}} \!\!=\!\! \left\{\begin{array}{ll} \!\!\eta_{_{L}}, \!\!\!&\text{if}~\theta_{xy} \!>\! \theta_c,\\ \!\!\eta_{_{N}}, \!\!\!&\text{if}~\theta_{xy} \!\le\! \theta_c, \end{array} \right.$
$\alpha_{_{xy}}$ denotes the path-loss exponent with $\alpha_{_{xy}} \!\!=\!\! \left\{\begin{array}{ll} \!\!\alpha_{_{L}}, \!\!\!&\text{if}~\theta_{xy} \!>\! \theta_c,\\ \!\!\alpha_{_{N}}, \!\!\!&\text{if}~\theta_{xy} \!\le\! \theta_c, \end{array} \right.$
and $S_{xy}$ denotes the channel power fading gain with $S_{xy} \!\!=\!\! \left\{\begin{array}{ll} \!\!1, \!\!\!&\text{if}~\theta_{xy} \!>\! \theta_c,\\\!\!S_N, \!\!\!&\text{if}~\theta_{xy} \!\le\! \theta_c. \end{array} \right.$
It is worth noting that the considered elevation-angle-dependent LoS/NLoS path-loss model is consistent with the practically measured dual-slope path-loss model in  \cite[(4)]{KhawajaSurvey2019}.

Furthermore, due to the strong LoS channel power gains over A2G links, we assume that the UAV-enabled wireless network is interference-limited by considering the UAVs are densely deployed, such that the background noise is negligible at each GN receiver. Under such assumptions, the mutual information or Shannon capacity of an A2G wireless link is determined by the received signal-to-interference ratio (SIR) at the GN. First, consider a typical legitimate receiver located at the origin of the ground plane, for which the received SIR is
\begin{align}
{{\tt SIR}_0} = \frac{{{\eta_{_L}}{P_t}}{{{H}^{{-2} }}}}{\sum\limits_{u \in {\Phi_{u}}} {{\eta_{_{u0}}}{{{P_t}{S_{u0}}}}{{{D_{u0}^{-\alpha_{u0}} }}}}}
= \frac{{{\eta_{_L}}}{{{H}^{{-2} }}}}{\sum\limits_{u \in {\Phi_{u}}} {{\eta_{_{u0}}}{{{S_{u0}}}}{{{D_{u0}^{-\alpha_{u0}} }}}}},
\label{sir_legitimate}
\end{align}
where ${S_{{u0}}}$ denotes the channel power fading gain from interfering UAV $u \in \Phi_u$ to the typical legitimate receiver, ${\eta_{_{u0}}}$ denotes the reference path loss of this link, ${\alpha_{u0}}$ denotes the path-loss exponent of this link, and ${D_{{u0}}}$ denotes their distance.
Note that according to the Slivnyak's Theorem \cite{Chiu2013}, it follows that the spatial distribution of the interfering UAVs still follows a homogeneous PPP with density ${\Phi_{u}}$, provided that the typical UAV transmitter's location is given. By slight abuse of notation, we denote ${\Phi_{u}}$ as the set of interfering UAVs in this paper.

Next, we consider any one particular eavesdropper $e\in \Phi_e$. Suppose that the eavesdroppers are non-cooperative and decode the typical UAV transmitter's sent messages individually. In this case, the received SIR at eavesdropper $e$ is
\begin{align}
{{\tt SIR}_e} = \frac{{\eta_{_{0e}}}{{P_t}}{S_{0e}}{{D_{0e}^{-{\alpha_{0e}}} }}}{\sum\limits_{u \in {\Phi_{u}}} {{\eta_{_{ue}}}{{{P_t}{S_{ue}}}}{{{D_{ue}^{-{\alpha_{ue}}} }}}}}
= \frac{{\eta_{_{0e}}}{S_{0e}}{{D_{0e}^{-{\alpha_{0e}}} }}}{\sum\limits_{u \in {\Phi_{u}}} {{\eta_{_{ue}}}{{{S_{ue}}}}{{{D_{ue}^{-{\alpha_{ue}}} }}}}},
\label{sir_eavesdropper}
\end{align}
where ${S_{{0e}}}$ denotes the channel power fading gain from the typical UAV transmitter to eavesdropper $e \in {\Phi _{e}}$, ${\eta_{_{0e}}}$ denotes the reference path loss of this link, ${\alpha_{0e}}$ denotes the path-loss exponent of this link, and ${D_{{0e}}}$ denotes their distance. Furthermore, ${S_{{ue}}}$ denotes the channel power fading gain from interfering UAV $u \in \Phi_u$ to eavesdropper $e \in {\Phi _{e}}$, ${\eta_{_{ue}}}$ denotes the reference path loss of this link, ${\alpha_{ue}}$ denotes the path-loss exponent of this link, and ${D_{{ue}}}$ denotes their distance.

Accordingly, the Shannon capacity of the legitimate channel between the typical UAV-to-legitimate-receiver pair (in bits-per-second-per-Hertz (bps/Hz)) is expressed as
\begin{align}
{C_t} = {{\log }_2}\left( {1 + {\tt SIR}_0} \right).
\label{capacity_legitimate}
\end{align}
The Shannon capacity of the eavesdropping channel from the typical UAV transmitter to the eavesdropper with the strongest SIR is expressed as
\begin{align}
{C_e} = {{\log }_2}\left( {1 + {\max\limits_{e \in {\Phi _{e}}}{\left\{ {\tt SIR}_{e}\right\} }}}\right).
\label{capacity_eavesdropper}
\end{align}
Accordingly, the achievable secrecy rate at the typical legitimate receiver is expressed as
 \begin{align}
 C_b = \left[C_t-C_e\right]^ + ,
 \end{align}
where ${\left[ x \right]^ + } \triangleq \max \left( {x,0} \right)$.
Here, notice that the secrecy rate is limited by the eavesdropper with the strongest SIR.

We adopt the well-established Wyner's encoding scheme for the UAVs' secrecy transmission \cite{Wyner1975}.
The Wyner's encoding scheme is implemented as follows in a nested nature based on two coding rates, namely the rate $R_t$ of the transmitted codewords and $R_s$ of the confidential information, where $R_t \ge R_s$ must hold. In particular, a wiretap code is constructed by generating $2^{M R_t}$ codewords $x^M(w,v)$ each of length $M$, where $w \in \{1,\ldots,2^{M R_s}\}$ and $v \in \{1,\ldots,2^{M (R_t-R_s)}\}$.
For each secrecy message with index $w$ to be secretly transmitted, we have several possible codewords $\{1,\ldots,2^{M (R_t-R_s)}\}$ that form a bin, which is named as a subcode of the wiretap code.
Accordingly, to transmit secrecy message with index $w$, we must randomly choose one codeword with index $v$ from these codewords $\{1,\ldots,2^{M (R_t-R_s)}\}$ with uniform probability to confuse eavesdroppers, and thus we have the codeword $x^M(w,v)$ to be transmitted.
Notice that the same values of $R_t$ and $R_s$ are used at different UAVs over the whole network, which are design variables to be optimized later. Accordingly, the rate difference ${R_e} = {R_t} - {R_s}$ represents the rate loss for transmitting the message securely against eavesdropping \cite{zhou2013rethinking}.

With the Wyner's encoding scheme, the typical UAV's data transmission is successful only when the Shannon capacity $C_t$ of the legitimate channel is larger than the rate ${R_t}$ of the transmitted codewords, i.e., the received ${{\tt SIR}_0}$ at the legitimate receiver exceeds the threshold ${\beta _t} = {2^{{R_t}}} - 1$. Then the connection probability, defined as the probability that the typical legitimate receiver is able to decode the UAV's transmitted message, is expressed as
\begin{align}
\nonumber{\mathcal{P}_{\textrm{c}}} &= {\mathcal{P}}\left( {C_t>R_t} \right)
= {\mathcal{P}}\left( {{\tt SIR}_0>{\beta _t}} \right)\\
&= {\mathcal{P}}\left( {\frac{{\eta_{_L}}{{{H}^{{-2} }}}}{\sum\limits_{u \in {\Phi_{u}}} {{\eta_{_0}}{{{S_{u0}}}}{{{D_{u0}^{-\alpha_{u0}} }}}}}>{\beta _t}} \right).
\label{P_c_1}
\end{align}
Furthermore, the typical UAV's transmitted confidential information can be successfully decoded at the eavesdroppers, when the Shannon capacity $C_e$ of the best eavesdropping link is larger than the rate difference ${R_e} = {R_t} - {R_s}$, i.e., the strongest SIR ${\max\limits_{e \in {\Phi _{e}}}{\left\{ {\tt SIR}_{e}\right\} }}$ exceeds the threshold ${\beta _e} = {2^{{R_e}}} - 1$. In this case, the secrecy outage occurs. Accordingly, the secrecy outage probability, defined as the probability that the eavesdropper is able to intercept the UAV's transmitted confidential information, is expressed as \cite{zhou2013rethinking}
\begin{align}
\nonumber{\mathcal{P}_{\textrm{so}}} &= {\mathcal{P}}\left (C_{\textrm{e}}> R_e\right)
= {\mathcal{P}}\left ({\max\limits_{e \in {\Phi _{e}}}{\left\{ {\tt SIR}_{e}\right\} }} > \beta_e\right)\\
&= {\mathcal{P}}\left ({\max\limits_{e \in {\Phi _{e}}}{\left\{ \frac{{\eta_{_0}}{S_e}{{D_{e}^{-{\alpha_{s}}} }}}{\sum\limits_{u \in {\Phi_{u}}} {{\eta_{_0}}{{{S_{ue}}}}{{{D_{ue}^{-{\alpha_{ue}}} }}}}}\right\} }} > \beta_e\right).
\label{P_so_1}
\end{align}
Accordingly, the secrecy transmission capacity of the UAV-enabled wireless network is defined as the average achievable rate of confidential messages that are successfully transmitted per unit area \cite{Zhou2011}, which is given as
\begin{align}
 \hat{\mathbb{C}}_s = {R_s}{{\mathcal{P}_c}}{\lambda_u}.
\label{capacity_1}
\end{align}

We are particularly interested in designing the system parameters (e.g., the wiretap code rates $R_t$ and $R_s$, and the UAVs' positioning altitude $H$), for the purpose of maximizing the secrecy transmission capacity $\hat{\mathbb{C}}_s$ of the UAV-enabled wireless network, while satisfying a maximum secrecy outage probability constraint. This, however, is generally a very challenging task, as the exact expressions of the secrecy communication performance metrics in (\ref{P_c_1}), (\ref{P_so_1}), and (\ref{capacity_1}) are not available yet.
Towards this end, in the following two sections we first derive the analytical expressions for these terms, and then solve the secrecy transmission capacity maximization problem.

\section{Secrecy Communication Performance Analysis}
In this section, we analyze the connection probability and secrecy outage probability of the typical UAV-and-legitimate-receiver pair. The analytical results will be used to measure the secrecy transmission capacity of the UAV-enabled wireless network later.

To facilitate the derivation, we denote $\Phi_{Ly}$ as the set of interfering UAVs with LoS wireless links to GN $y$, i.e., for any interfering UAV $l \in \Phi_{Ly}$, its location $\mv w_l \in \mathbb{R}^{3\times 1}$ satisfies $\|\mv w_l - \mv w_y\| \le \sqrt{  K^2+H^2 }$. Then we have the following two theorems.

\begin{theorem}\label{Theorem_P_c}
The connection probability of a typical UAV-and-legitimate-receiver pair ${\mathcal{P}_{\textrm{c}}}$ in (\ref{P_c_1}) is given by
\begin{align}
\nonumber{\mathcal{P}_{\textrm{c}}} = {\mathbb{E}_{\Phi_{u}}}\Bigg\{\sum\limits_{u \in {\Phi_{u}}\setminus{\Phi_{L0}}}\prod\limits_{{m \in {\Phi_{u}}\setminus{\Phi_{L0}}},m \ne u} {\frac{{{D_{m0}^{\alpha_{_N}} }}}{{{{{D_{m0}^{\alpha_{_N}} }}}  - {{{D_{u0}^{\alpha_{_N}} }}} }}} \\
\times\left(1-\exp \left[ { - \frac{{\eta_{_L}}{D_{u0}^{\alpha_{_N}} }}{{\eta_{_N}}{\beta_t}H^2}+\frac{\eta_{_L}{{D_{u0}^{\alpha_{_N}} }}}{\eta_{_N}}{\sum\limits_{l \in {\Phi_{L0}}} {{{{D_{l0}^{-{\alpha_{_L}}} }}}}}   } \right]\right)\Bigg\}.
\label{P_c_exact}
\end{align}

\begin{IEEEproof}
See Appendix \ref{appendices_Theorem_P_c}.
\end{IEEEproof}
\end{theorem}
\vspace{3ex}

\begin{theorem}\label{Theorem_P_so_no_zone}
The secrecy outage probability of a typical UAV-to-legitimate-receiver pair ${\mathcal{P}_{\textrm{so}}}$ in (\ref{P_so_1}) is given by
\begin{align}
\label{P_so_exact}&{\mathcal{P}_{\textrm{so}}}=1-{\mathbb{E}_{\Phi_{u}}}\Bigg\{\exp\Bigg[ { - {\lambda_{{e}}}} \Bigg\{\int_{\boldsymbol{B}\left ( o, K \right )}\\
\nonumber&\sum\limits_{u \in {\Phi_{u}}\setminus{\Phi_{Le}}}{\prod\limits_{{m \in {\Phi_{u}}\setminus{\Phi_{Le}}},m \ne u} {\frac{{{D_{me}^{\alpha_{_N}} }}}{{{{{D_{me}^{\alpha_{_N}} }}}  - {{{D_{ue}^{\alpha_{_N}} }}} }}} }\\
\nonumber&\left(1-\exp \left[ { - \frac{{\eta_{_L}}{D_{ue}^{\alpha_{_N}} }}{{\eta_{_N}}{\beta_e}{{D_{0e}^{\alpha_{_L}} }}}+\frac{\eta_{_L}{{D_{ue}^{\alpha_{_N}} }}}{\eta_{_N}}{\sum\limits_{l \in {\Phi_{Le}}} {D_{le}^{-\alpha_{_L} }}}   } \right]\right)\Bigg\}\, \text{d}e\\
\nonumber&+\int_{{\mathbb{R}^2\setminus{\boldsymbol{B}\left ( o, K \right )}}} {\prod\limits_{{l} \in {\Phi_{Le}} } {\left\{{\exp\left [-{\frac{{\eta_{_L}}{\beta_e}{{D_{0e}^{\alpha_{_N}} }}}{{\eta_{_N}} {{{{D_{le}^{\alpha_{_L}} }}}}}} \right]}\right\}} } \\
\nonumber&\times{\prod\limits_{{u} \in {\Phi_{u}}\setminus{\Phi_{Le}} } {\left\{ \frac{1}{1+{{\beta_e}{{D_{0e}^{\alpha_{_N}} }}{{D_{ue}^{-{\alpha_{_N}}} }}}}\right\}} } \, \text{d}e \Bigg\}\Bigg]\Bigg\},
\end{align}
where $\boldsymbol{B}\left ( o, K \right )$ represents a disk of radius $K$ centered at $o$.
\begin{IEEEproof}
See Appendix \ref{appendices_Theorem_P_so_no_zone}.
\end{IEEEproof}
\end{theorem}
\vspace{3ex}

Theorems \ref{Theorem_P_c} and \ref{Theorem_P_so_no_zone} show the exact expressions of the connection probability and secrecy outage probability for the UAV-enabled wireless network. However, due to the involvement of statistical expectation in (\ref{P_c_exact}) and (\ref{P_so_exact}), these expressions are too complicated to draw insights. Besides, it is difficult to solve the secrecy transmission capacity maximization problem based on these expressions. To resolve these issues, in the following we resort to approximate expressions of the connection probability and secrecy outage probability. Notice that the key difficulty in deriving mathematically tractable connection and secrecy outage probabilities lies in the fact that the A2G channels are LoS when the elevation angle exceeds threshold $\theta_c$.
In this case, the A2G channel power gain and accordingly the received signal power in the SIR expressions become constant, under which it is very difficult to handle the expectation over SIR terms as the received interference power corresponds to the sum of various independent exponential random variables. This thus leads to complicated expressions in (\ref{P_c_exact}) and (\ref{P_so_exact}) with statistical expectation involved. By contrast, if the A2G channel is NLoS when elevation angle is no larger than $\theta_c$, the channel power gain and the received signal power in the SIR terms become exponentially random variables. By exploiting this feature, the expectation over the received interference power (i.e., sum of independent exponential random variables) can be turned into a tractable form.
Motivated by this fact, we propose to approximate the LoS A2G channel into a Rayleigh fading channel with the same path loss.
In other words, for any A2G channel from UAV $x$ to GN $y$, we consider that the channel fading power gain $S_{xy}$ is an exponentially distributed random variable with unit mean, denoted by $\bar{S}_{xy}$, regardless of the elevation angle.
Accordingly, the received SIR at the typical legitimate receiver and that at eavesdropper $e$ are respective approximated as
\begin{align}
{\overline{\tt SIR}_0} = \frac{{{\eta_{_L}}{P_t}{\overline{S}_{00}}}{{{H}^{{-2} }}}}{\sum\limits_{u \in {\Phi_{u}}} {{{\eta_{_{u0}}}{{P_t}{\overline{S}_{u0}}}}{{{D_{u0}^{-\alpha_{u0}} }}}}}
= \frac{{{\eta_{_L}}{\overline{S}_{00}}}{{{H}^{{-2} }}}}{\sum\limits_{u \in {\Phi_{u}}} {{{\eta_{_{u0}}}{{\overline{S}_{u0}}}}{{{D_{u0}^{-\alpha_{u0}} }}}}},
\label{sir_legitimate_fading}
\end{align}

\begin{align}
{\overline{\tt SIR}_e} = \frac{{{\eta_{_{0e}}}{P_t}{\overline{S}_{0e}}}{{D_{0e}^{-\alpha_{0e}} }}}{\sum\limits_{u \in {\Phi_{u}}} {{{{\eta_{_{ue}}}{P_t}{\overline{S}_{ue}}}}{{{D_{ue}^{-\alpha_{ue}} }}}}}
= \frac{{{\eta_{_{0e}}}{\overline{S}_{0e}}}{{D_{0e}^{-\alpha_{0e}} }}}{\sum\limits_{u \in {\Phi_{u}}} {{{{\eta_{_{ue}}}{\overline{S}_{ue}}}}{{{D_{ue}^{-\alpha_{ue}} }}}}}.
\label{sir_eavesdropper_fading}
\end{align}
Accordingly, the approximate connection and secrecy outage probabilities of a typical UAV-and-legitimate-receiver pair are respectively denoted as
\begin{align}
{\tilde{\mathcal{P}}_{\textrm{c}}} &= {\mathcal{P}}\left (\frac{{{\eta_{_L}}{\overline{S}_{00}}}{{{H}^{{-2} }}}}{\sum\limits_{u \in {\Phi_{u}}} {{{{\eta_{_{u0}}}{\overline{S}_{u0}}}}{{{D_{u0}^{-\alpha_{u0}} }}}}} > \beta_t \right),
\label{P_c_1_approx}
\end{align}
\begin{align}
{\tilde{\mathcal{P}}_{\textrm{so}}}=& {\mathcal{P}}\left ({\mathop {\max }\limits_{{e} \in {\Phi _{e}}} \left\{\frac{{{\eta_{_{0e}}}{\overline{S}_{0e}}}{{D_{0e}^{-\alpha_{0e}} }}}{\sum\limits_{u \in {\Phi_{u}}} {{{{\eta_{_{ue}}}{\overline{S}_{ue}}}}{{{D_{ue}^{-\alpha_{ue}} }}}}} \right\}}> \beta_e \right).
\label{P_so_0_no_zone}
\end{align}
Then we have the following two theorems.

\begin{corollary}\label{Corollary_P_c_approx}
The approximate connection probability of a typical UAV-and-legitimate-receiver pair in (\ref{P_c_1_approx}) is given by
\begin{align}
\nonumber\tilde{\mathcal{P}}_{{\rm{c}}}=& \exp \Bigg[  - {{\frac{{\pi}{\lambda_u}{H}{\beta_t^{\frac{1}{2}}}{\eta_{_N}^{\frac{1}{2}}}}{2{\eta_{_L}^{\frac{1}{2}}}}}}\\
\nonumber&\times\left(\pi-2\arctan\left(\frac{{\eta_{_L}^{\frac{1}{2}}}\left({H}^{2}+{K}^{2}\right)}{{\eta_{_N}^{\frac{1}{2}}}{H}{\beta_t^{\frac{1}{2}}}}\right) \right)\\
&-{\pi}{\lambda_u}{H}^{2}{\beta_t}\log\left(\frac{{H}^{2}{\beta_t}+{H}^{2}+{K}^{2}}{{H}^{2}{\beta_t}+{H}^{2}} \right)\Bigg],
\label{P_c_final}
\end{align}

\begin{IEEEproof}
See Appendix \ref{appendices_Corollary_P_c_approx}.
\end{IEEEproof}
\end{corollary}
\vspace{3ex}

\begin{corollary}\label{Corollary_P_so_no_zone_approx}
The approximate secrecy outage probability of a typical UAV-and-legitimate-receiver pair in (\ref{P_so_0_no_zone}) is given by
\begin{align}
\nonumber&\tilde{\mathcal{P}}_{{\textrm{so}}} = 1 - \exp \Bigg[{ - 2\pi\lambda_e }\exp \left[{\pi }{\lambda_u}{H^2}\right]\\
\nonumber&\times \Bigg\{{ \frac{{\eta_{_L}^{\frac{1}{2}}}}{2{\eta_{_N}^{\frac{1}{2}}}{Q_1}}}\exp \left[-\frac{\eta_{_N}^{\frac{1}{2}}}{{\eta_{_L}^{\frac{1}{2}}}}{{Q_1}\left ( {H^2}+{K^2}\right )}\right]\\
\nonumber&+\frac{\left ( 1+H{Q_1}\right )\exp \left[-H{Q_1}\right]}{Q_1^2}\\
&-\frac{{\left ( 1+{Q_1}\sqrt{H^2+K^2}\right )\exp \left[-{Q_1}\sqrt{H^2+K^2}\right]}}{Q_1^2}
\Bigg\}\Bigg],
\label{P_so_final_no_zone}
\end{align}
where $Q_1 = \frac{{{\lambda_u}\pi^2\beta_e^{\frac{1}{2}}}}{2}$.
\begin{IEEEproof}
See Appendix \ref{appendices_Corollary_P_so_no_zone_approx}.
\end{IEEEproof}
\end{corollary}
\vspace{3ex}

Corollaries \ref{Corollary_P_c_approx} and \ref{Corollary_P_so_no_zone_approx} provide the connection probability and secrecy outage probability in closed forms, which can provide insights and facilitate the system design later in Section \uppercase\expandafter{\romannumeral4}. For example, Corollary \ref{Corollary_P_c_approx} shows that the approximate connection probability $\tilde{\mathcal{P}}_{{\rm{c}}}$ is monotonically decreasing with respect to the radius $K$ of the LoS region. This is due to the fact that as $K$ increases, more interfering links are strong LoS links, thus leading to stronger interference and reduced $\tilde{\mathcal{P}}_c$.

\begin{figure}[!h]
\centering
  \includegraphics[width=6.8cm]{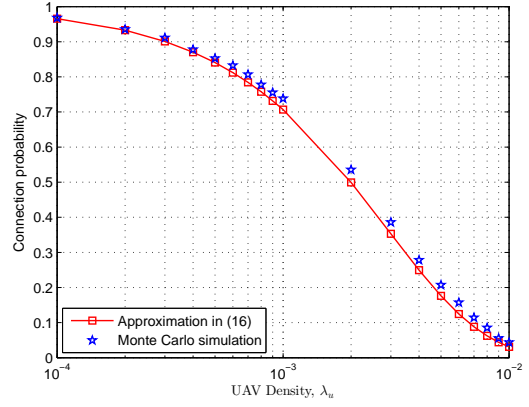}\\
  \caption{The connection probability versus the UAVs' (or legitimate receivers') density $\lambda_u$, where we set $R_t = 5$, and ${\theta_{c}} = \pi/4$.}
\label{fig:fig_P_c}
\end{figure}

\begin{figure}[!h]
\centering
  \includegraphics[width=6.8cm]{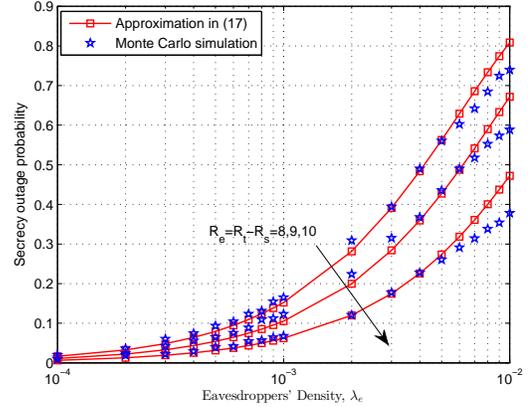}\\
  \caption{The secrecy outage probability versus the eavesdroppers' density $\lambda_e$, where we set $\lambda_u = 10^{-3}$, $H = 10$~m, and ${\theta_{c}} = \pi/4$.}
\label{fig:fig_P_so}
\end{figure}

Figs. \ref{fig:fig_P_c} and \ref{fig:fig_P_so} show the approximate connection probability in (\ref{P_c_final}) and secrecy outage probability in (\ref{P_so_final_no_zone}), respectively, as compared to the Monte Carlo simulations. It is observed in Fig. \ref{fig:fig_P_c} that our proposed approximate connection probability in (\ref{P_c_final}) matches well with the Monte Carlo simulation results.
Furthermore, it is observed in Fig. \ref{fig:fig_P_so} that our proposed approximate secrecy outage probability in (\ref{P_so_final_no_zone}) matches well with the Monte Carlo simulation results under practical scenarios when the secrecy outage probability is small (e.g., $0\sim0.1$).
This validates the accuracy of approximating the A2G LoS channels into Rayleigh fading channels when the elevation angles are smaller than $\theta_c$. Therefore, the approximation into Rayleigh fading channels is not only tractable but also practically reasonable under our consideration. Notice that the Nakagami-m or Rician fading may be more accurate to model or approximate the LoS-dominated A2G channels, which can be left for future work.

\section{Secrecy Transmission Capacity Maximization}
In this section, based on the above approximations, we jointly design the wiretap code rates $R_t$ and $R_s$, and the UAV positioning altitude $H$ to maximize the secrecy transmission capacity while satisfying a maximum secrecy outage probability constraint. With the approximations in (\ref{P_c_final}) and (\ref{P_so_final_no_zone}), the secrecy maximization problem is expressed as
\begin{subequations}
\begin{align}
\label{P1_a}(\textbf{P1.1}):~~ \underset{R_t,R_s,H}{\max}\quad &\tilde{\mathbb{C}}_s = {R_s}{{\tilde{\mathcal{P}}_c}}{\lambda_u}, \\
\label{P1_b}\mathrm{s.t.}\quad &{\tilde{\mathcal{P}}_{\textrm{so}}}\leq \epsilon,\\
\label{P1_c}&R_t \geq R_s,\\
\label{P1_d}&R_s \geq 0,\\
\label{P1_e}&H_\textrm{max} \geq H \geq H_\textrm{min},
\end{align}
\end{subequations}
where $\epsilon\in \left[0,1\right]$ represents the minimum security requirement.

Notice that problem ($\textbf{P1.1}$) is still difficult to be optimally solved, as the design parameters are highly interrelated. In the following, we first obtain the closed-form solution of $R_t$ and $R_s$ under any given UAV positioning altitude $H$, and then use a one-dimensional (1D) search over $H \in [H_\textrm{min}, H_\textrm{max}]$ to maximize the secrecy transmission capacity. In the following, we only need to focus on finding the optimal $R_t$ and $R_s$ under any given $H$, for which the problem is expressed as
\begin{subequations}
\begin{align}
\label{P2_a}(\textbf{P1.2}):~ \underset{R_t,R_s}{\max}\quad &\tilde{\mathbb{C}}_s = {{ {R_s}}}{{\tilde{\mathcal{P}}_c}}{\lambda_u}, \\
\label{P2_b}\mathrm{s.t.}\quad &{\tilde{\mathcal{P}}_{\textrm{so}}}\leq \epsilon,\\
\label{P2_c} &R_t \geq R_s,\\
\label{P2_d} &R_s \geq 0.
\end{align}
\end{subequations}

In order to solve problem ($\textbf{P1.2}$), notice that ${\tilde{\mathcal{P}}_{\textrm{so}}}$ in (\ref{P2_b}) only depends on $R_t - R_s$. To facilitate the derivation, we introduce the auxiliary variable $R_e = R_t - R_s$. By substituting $R_s = R_t - R_e$, problem ($\textbf{P1.2}$) is equivalent to
\begin{subequations}
\begin{align}
\label{P3_a}(\textbf{P1.3}):~ \underset{R_t,R_e}{\max}\quad &\tilde{\mathbb{C}}_s = {{\left ( {R_t - R_e} \right )}}{{\tilde{\mathcal{P}}_c}}{\lambda_u}, \\
\label{P3_b}\mathrm{s.t.}\quad &{\tilde{\mathcal{P}}_{\textrm{so}}}\leq \epsilon,\\
\label{P3_c} &R_t \geq R_e,\\
\label{P3_d} &R_e \geq 0.
\end{align}
\end{subequations}

Notice that from (\ref{P_so_0_no_zone}), it follows that $\tilde{P}_{so}$ is monotonically decreasing with respect to $\beta_e $, where ${\beta _e} = {2^{{R_e}}} - 1 $. Therefore, it is evident that both $\tilde{P}_{so}$ and $\tilde{\mathbb{C}}_s$ are monotonically decreasing functions with respect to $R_e \ge 0$. Hence, at the optimality of problem ($\textbf{P1.3}$), constraint (\ref{P3_b}) must be met with strict equality, i.e.,
\begin{align}
{\tilde{\mathcal{P}}_{\textrm{so}}} = \epsilon.
\label{P_so_no_zone_R_e}
\end{align}
By using a 1D search to get a unique solution to the equality in (\ref{P_so_no_zone_R_e}), which is the optimal solution of $R_e$ to problem ($\textbf{P1.3}$), denoted by $R_e^*$.

Next, it remains to find the solution of $R_t$ to problem ($\textbf{P1.3}$). Notice that for $x \to 0$, it holds that $\arctan\left(x\right) \approx x$ and $\log(1+x) \approx x$. As $\beta_t \gg 1$ generally holds, we have
\begin{align}
{\tilde{\mathcal{P}}_{{\rm{c}}} \approx \bar{\mathcal{P}}_{{\rm{c}}} = \exp \left[ { - {{\frac{{\pi}}{2}}}{\lambda_u}{H}\left({\frac{{\eta_{_N}^{\frac{1}{2}}}}{{\eta_{_L}^{\frac{1}{2}}}}}{\pi}{2^{\frac{R_t}{2}}}-2{H} \right)} \right]}.
\label{P_c_approx}
\end{align}

In this case, solving for $R_t$ in problem ($\textbf{P1.3}$) is approximated as
\begin{subequations}
\begin{align}
\label{P32_a}(\textbf{P1.4}):~ \underset{R_t}{\max}\quad &\mathbb{C}_s = {{\left ( {R_t - R_e^*} \right )}}{{\bar{\mathcal{P}}_c}}{\lambda_u}, \\
\label{P32_b}\mathrm{s.t.}\quad &R_t \geq R_e^*.
\end{align}
\end{subequations}
We then have the following theorem.
\begin{theorem}\label{Theorem_R_t_R_s}
The objective function $\mathbb{C}_s$ in (\ref{P32_a}) is a concave function with respect to $R_t \ge 0$. The optimal solution of $R_t$ to problem ($\textbf{P1.4}$) is given as
\begin{align}
R_t^* = R_e^* + \frac{2}{\ln2}\mathbb{W}_0\left (\frac{{\eta_{_L}^{\frac{1}{2}}}2^{-R_e^*+1}}{{\eta_{_N}^{\frac{1}{2}}}{\pi^2}{\lambda_u}{H}} \right ).
\label{R_t_no_zone}
\end{align}
\begin{IEEEproof}
See Appendix \ref{appendices_Theorem_R_t_R_s}.
\end{IEEEproof}
\end{theorem}
\vspace{3ex}
By combining $R_t^*$ and $R_e^*$, the corresponding $R_s$ is obtained as
\begin{align}
R_s^* = \frac{2}{\ln2}\mathbb{W}_0\left (\frac{{\eta_{_L}^{\frac{1}{2}}}2^{-R_e^*+1}}{{\eta_{_N}^{\frac{1}{2}}}{\pi^2}{\lambda_u}{H}} \right ).
\label{R_s_no_zone}
\end{align}
Therefore, by combining $R_t^*$ and $R_s^*$, together with the 1D search over $H$, the desirable system configurations are finally obtained.

\section{Secrecy Guard Zone}
In this section, we adopt the secrecy guard zone technique to further protect the UAVs' secrecy transmission. Although the guard zone technique has been widely applied in conventional secrecy wireless communications \cite{Zhou2011,PintoSecure2012,KoyluogluOn2012,XuSecure2016,YangDelivery2018,TangImpact2018}, the employment in secrecy UAV communications is new and has not been studied yet.
In particular, we suppose that each UAV is able to physically detect the existence of suspicious eavesdroppers within a certain finite region, namely the secrecy guard zone \cite{Zhou2011}.
This region is modeled as a disk on the ground, with a given radius $D$ centered at the UAV's horizontal projection.
Accordingly, each UAV transmits confidential messages only when there is no eavesdropper inside the corresponding secrecy guard zone.
If there are any eavesdropper(s) found inside the secrecy guard zone, the UAV will generate artificial noise (AN) to confuse these eavesdroppers for helping mask nearby UAVs' confidential message transmissions. Note that the AN is assumed to be statistically identical to the confidential messages and hence cannot be distinguished by eavesdroppers. Also note that this cooperative protocol is implemented in a distributed manner, and does not require any coordination between UAV transmitters.

\subsection{Secrecy Communication Performance Analysis}
In this subsection, we analyze the connection probability and secrecy outage probability of the typical UAV-and-legitimate-receiver pair, in the case with the guard zone employed.

For any legitimate receiver or eavesdropper, the set of interfering UAVs can still be denoted as ${\Phi_{u}}$ with density $\lambda_u$, which is same as that in Section \uppercase\expandafter{\romannumeral3} without the secrecy guard zone used.
Hence, the exact and approximate connection probabilities are also given as $\mathcal{P}_{{\rm{c}}}$ in (\ref{P_c_exact}) and $\tilde{\mathcal{P}}_{{\rm{c}}}$ in (\ref{P_c_final}), respectively. Nevertheless, with the secrecy guard zone, the set of UAV transmitters is reduced as ${\Phi_{u}}'$ with density ${\lambda _u}'$ given by \cite{Zhou2011}
\begin{align}
{\lambda _u}' = {\lambda _u}\exp\left [-\pi{\lambda_e}{D^2}\right].
\label{transmitter_density}
\end{align}
Accordingly, for the typical legitimate receiver at the origin, the set of eavesdroppers ${\Phi _{e}}$ follows a homogeneous PPP with density ${\lambda_e}$ outside $\boldsymbol{B}\left (o,D \right )$.
In this case, by modifying the distributed area of the eavesdroppers from the whole ground plane ${\mathbb{R}^2}$ in (\ref{P_so_1}) to ${{\mathbb{R}^2}\setminus{\boldsymbol{B}\left ( o, D \right )}}$, the secrecy outage probability ${\mathcal{P}_{\textrm{so}}}$ is re-expressed as
\begin{align}
{\mathcal{P}_{\textrm{so}}^{\textrm{zone}}} =  {\mathcal{P}}\left ({\max\limits_{\|\mv w_e\| \geq D,\forall e \in {\Phi _{e}}}{\left\{ \frac{{\eta_{_{0e}}}{S_{0e}}{{D_{0e}^{-{\alpha_{0e}}} }}}{\sum\limits_{u \in {\Phi_{u}}} {{\eta_{_{ue}}}{{{S_{ue}}}}{{{D_{ue}^{-{\alpha_{ue}}} }}}}}\right\} }} > \beta_e\right).
\label{P_so_zone_1}
\end{align}
We have the following theorem.

\begin{theorem}\label{Theorem_P_so}
Under the secrecy guard zone with radius $D$, the secrecy outage probability of a typical UAV-and-legitimate-receiver pair is given by
\begin{align}\label{P_so_exact_zone}
\nonumber&{\mathcal{P}_{\textrm{so}}^{\textrm{zone}}}=1-{\mathbb{E}_{\Phi_{u}}}\Bigg\{\exp\Bigg[ { - {\lambda_{{e}}}} \Bigg\{\int_{\boldsymbol{B}\left ( o, K \right )\setminus{\boldsymbol{B}\left ( o, D\right )}} \sum\limits_{u \in {\Phi_{u}}\setminus{\Phi_{Le}}}\\
\nonumber&{\prod\limits_{{m \in {\Phi_{u}}\setminus{\Phi_{Le}}},m \ne u} {\frac{{{D_{me}^{\alpha_{_N}} }}}{{{{{D_{me}^{\alpha_{_N}} }}}  - {{{D_{ue}^{\alpha_{_N}} }}} }}} }\\
\nonumber&\left(1-\exp \left[ { - \frac{{\eta_{_L}}{D_{ue}^{\alpha_{_N}} }}{{\eta_{_N}}{\beta_e}{{D_{0e}^{\alpha_{_L}} }}}+\frac{\eta_{_L}{{D_{ue}^{\alpha_{_N}} }}}{\eta_{_N}}{\sum\limits_{l \in {\Phi_{Le}}} {D_{le}^{-\alpha_{_L} }}}   } \right]\right)\Bigg\}\, \text{d}e\\
\nonumber&+\int_{{\mathbb{R}^2\setminus{\boldsymbol{B}\left ( o, K \right )}}} {\prod\limits_{{l} \in {\Phi_{Le}} } {\left\{{\exp\left [-{\frac{{\eta_{_L}}{\beta_e}{{D_{0e}^{\alpha_{_N}} }}}{{\eta_{_N}} {{{{D_{le}^{\alpha_{_L}} }}}}}} \right]}\right\}} } \\
&\times{\prod\limits_{{u} \in {\Phi_{u}}\setminus{\Phi_{Le}} } {\left\{ \frac{1}{1+{{\beta_e}{{D_{0e}^{\alpha_{_N}} }}{{D_{ue}^{-{\alpha_{_N}}} }}}}\right\}} } \, \text{d}e \Bigg\}\Bigg]\Bigg\}.
\end{align}
\begin{IEEEproof}
See Appendix \ref{appendices_Theorem_P_so}.
\end{IEEEproof}
\end{theorem}
\vspace{3ex}

Notice that the exact expression of the secrecy outage probability in Theorem \ref{Theorem_P_so} is too complicated. Therefore, similarly as in Section \uppercase\expandafter{\romannumeral3}, we approximate the LoS wireless channel power gain as a Rayleigh fading one with the same path loss and denote the correspondingly achieved SIRs as those in (\ref{sir_legitimate_fading}) and (\ref{sir_eavesdropper_fading}). Then we have the following theorem.

\begin{corollary}\label{Corollary_P_so_approx}
Under the secrecy guard zone with radius $D$, the approximate secrecy outage probability of a typical UAV-and-legitimate-receiver pair is given by
\begin{equation}
 {\tilde{\mathcal{P}}_{{\textrm{so}}}^{\textrm{zone}}} \!=\! \left\{\!\!\!
 \begin{array}{lllll}
1-&\!\!\!\!\!\exp \Bigg[ -\frac{{\pi}{\lambda_e}}{Q_1}
\exp \bigg[-{Q_1}&\\
&\times\left( {H^2}+{D^2}\right)+{\pi}{\lambda_u}{H^2}\bigg]\Bigg], &\!\!\!\!\!\!\text{if}~{D} \geq K,\\
1 - &\!\!\!\!\!\exp \Bigg[{ -2\pi\lambda_e }\exp \left[{\pi }{\lambda_u}{H^2}\right]&\\
\times&\!\!\!\!\!\!\!\!\Bigg\{{ \frac{{\eta_{_L}^{\frac{1}{2}}}}{2{\eta_{_N}^{\frac{1}{2}}}{Q_1}}}\exp \left[-\frac{\eta_{_N}^{\frac{1}{2}}}{{\eta_{_L}^{\frac{1}{2}}}}{{Q_1}\left ( {H^2}+{K^2}\right )}\right]\\
+&\!\!\!\!\!\!\!\frac{{\left ( 1+{Q_1}\sqrt{H^2+D^2}\right )\exp \left[-{Q_1}\sqrt{H^2+D^2}\right]}}{Q_1^2}&\\
-&\!\!\!\!\!\!\!\! \frac{{\left ( 1+{Q_1}\sqrt{H^2+K^2}\right )\exp \left[-{Q_1}\sqrt{H^2+K^2}\right]}}{Q_1^2}
\Bigg\}\Bigg],&\!\!\!\!\!\!\text{if}~{D} < K.
\end{array}
\right.
\label{P_so_final_zone}
\end{equation}
\begin{IEEEproof}
See Appendix \ref{appendices_Corollary_P_so_approx}.
\end{IEEEproof}
\end{corollary}
\vspace{3ex}

Corollary \ref{Corollary_P_so_approx} provides the secrecy outage probability in a closed form, which provide design insights. For example, it is shown that $\tilde{\mathcal{P}}_{\rm{so}}^{\textrm{zone}}$ is monotonically decreasing with respect to the secrecy guard zone radius $D$. This is expected, as a larger $D$ reduces the chance for eavesdropping.

\begin{figure}[!h]
\centering
  \includegraphics[width=6.8cm]{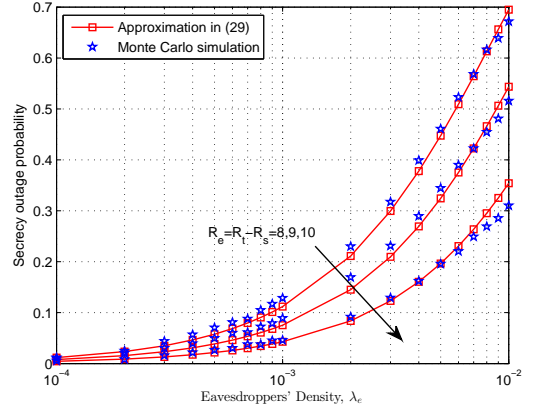}\\
  \caption{The secrecy outage probability with secrecy guard zone versus the eavesdroppers' density $\lambda_e$, where we set $\lambda_u = 10^{-3}$, $H = 10$~m, and ${\theta_{c}} = \pi/4$.}
\label{fig:fig_P_so_zone}
\end{figure}
Fig. \ref{fig:fig_P_so_zone} shows the approximate secrecy outage probability ${\tilde{\mathcal{P}}_{{\textrm{so}}}^{\textrm{zone}}}$ in (\ref{P_so_final_zone}), as compared to the Monte Carlo simulations. It is observed that our proposed approximation matches well with the Monte Carlo simulation results under practical scenarios when the secrecy outage probability is small (e.g., $0\sim0.1$).

\subsection{Secrecy Transmission Capacity Maximization}
In this subsection, based on the approximations, we jointly design the wiretap code rates $R_t$ and $R_s$, the UAV positioning altitude $H$, and the guard zone radius $D$ to maximize the secrecy transmission capacity while satisfying a maximum secrecy outage probability constraint. With the approximations in (\ref{P_c_final}) and (\ref{P_so_final_zone}), the secrecy maximization problem is expressed as
\begin{subequations}
\begin{align}
\label{P4_a}(\textbf{P2.1}):~~ \underset{R_t,R_s,H,D}{\max}\quad &\tilde{\mathbb{C}}_s = {R_s}{{\tilde{\mathcal{P}}_c}}{\lambda_u}', \\
\label{P4_b}\mathrm{s.t.}\quad &{\tilde{\mathcal{P}}_{{\textrm{so}}}^{\textrm{zone}}}\leq \epsilon,\\
\label{P4_c}&R_t \geq R_s,\\
\label{P4_d}&R_s \geq 0,\\
\label{P4_e}&D \geq 0,\\
\label{P4_f}&H_\textrm{max} \geq H \geq H_\textrm{min}.
\end{align}
\end{subequations}
It can be observed in (\ref{P4_a}) that a larger guard zone radius $D$ leads to lower secrecy outage probability ${\tilde{\mathcal{P}}_{{\textrm{so}}}^{\textrm{zone}}}$, but also reduces the density of actual UAV transmitters. As a result, there generally exists a tradeoff in designing the guard zone radius $D$.

Notice that problem ($\textbf{P2.1}$) is even more difficult to be optimally solved than ($\textbf{P1.1}$). In the following, we first obtain the closed-form solution of $R_t$ and $R_s$ under any given UAV positioning altitude $H$ and guard zone radius $D$, and then use a two-dimensional (2D) search over $H \in [H_\textrm{min}, H_\textrm{max}]$ and $D \geq 0$ to maximize the secrecy transmission capacity. In the following, we focus on finding the optimal $R_t$ and $R_s$ under any given $H$ and $D$, for which the problem is expressed as
\begin{subequations}
\begin{align}
\label{P5_a}(\textbf{P2.2}):~ \underset{R_t,R_s}{\max}\quad &\tilde{\mathbb{C}}_s = {{ {R_s}}}{{\tilde{\mathcal{P}}_c}}{\lambda_u}', \\
\label{P5_b}\mathrm{s.t.}\quad &{\tilde{\mathcal{P}}_{{\textrm{so}}}^{\textrm{zone}}}\leq \epsilon,\\
\label{P5_c} &R_t \geq R_s,\\
\label{P5_d} &R_s \geq 0.
\end{align}
\end{subequations}

In order to solve problem ($\textbf{P2.2}$), similarly as for ($\textbf{P1.2}$),  we introduce the auxiliary variable $R_e = R_t - R_s$. By substituting $R_s = R_t - R_e$, problem ($\textbf{P2.2}$) is equivalent to
\begin{subequations}
\begin{align}
\label{P6_a}(\textbf{P2.3}):~ \underset{R_t,R_e}{\max}\quad &\tilde{\mathbb{C}}_s = {{\left ( {R_t - R_e} \right )}}{{\tilde{\mathcal{P}}_c}}{\lambda_u}', \\
\label{P6_b}\mathrm{s.t.}\quad &{\tilde{\mathcal{P}}_{{\textrm{so}}}^{\textrm{zone}}}\leq \epsilon,\\
\label{P6_c} &R_t \geq R_e,\\
\label{P6_d} &R_e \geq 0.
\end{align}
\end{subequations}

Notice that both ${\tilde{\mathcal{P}}_{{\textrm{so}}}^{\textrm{zone}}}$ and $\tilde{\mathbb{C}}_s$ are monotonically decreasing functions with respect to $R_e \ge 0$. Hence, at the optimality of problem ($\textbf{P2.3}$), constraint (\ref{P6_b}) must be met with strict equality, i.e.,
\begin{align}
{\tilde{\mathcal{P}}_{{\textrm{so}}}^{\textrm{zone}}} = \epsilon.
\label{P_so_zone_R_e}
\end{align}
By solving the equality in (\ref{P_so_zone_R_e}) via a 1D search, we can obtain the optimal $R_e$, denoted by $R_e^\star$. Notice that in the special case with $D \geq L$, the optimal $R_e^\star$ can be obtained in closed-form as
\begin{align}
\nonumber R_e^\star &= \log_2\Bigg [{\frac{4}{{\pi^4}{\lambda_u}^2\left ({H^2+D^2}\right )^2}}\\
&\times \mathbb{W}_0\left (\frac{{{\pi}{\lambda_e}\left ({H^2+D^2}\right )}{\exp\left [ {{{\pi}{\lambda_u}{H^2}}}\right ]}}{{\ln{\frac{1}{1-\epsilon}}}}  \right )^2 +1 \Bigg ].
\label{R_e_zone}
\end{align}

Next, it remains to find $R_t$ to problem ($\textbf{P2.3}$). Similar to ($\textbf{P1.3}$), solving for $R_t$ in problem ($\textbf{P2.3}$) can be approximated as
\begin{subequations}
\begin{align}
\label{P7_a}(\textbf{P2.4}):~ \underset{R_t}{\max}\quad &\mathbb{C}_s = {{\left ( {R_t - R_e^\star} \right )}}{{\bar{\mathcal{P}}_c}}{\lambda_u}', \\
\label{P7_b}\mathrm{s.t.}\quad &R_t \geq R_e^*.
\end{align}
\end{subequations}
We then have the following theorem.
\begin{theorem}\label{Theorem_R_t_R_s_zone}
The objective function $\mathbb{C}_s$ in (\ref{P7_a}) is a concave function with respect to $R_t \ge 0$. The optimal solution of $R_t$ to problem ($\textbf{P2.4}$) is given as
\begin{align}
R_t^\star = R_e^\star + \frac{2}{\ln2}\mathbb{W}_0\left (\frac{{\eta_{_L}^{\frac{1}{2}}}2^{-R_e^\star+1}}{{\eta_{_N}^{\frac{1}{2}}}{\pi^2}{\lambda_u}{H}} \right ).
\label{R_t_zone}
\end{align}
\begin{IEEEproof}
See Appendix \ref{appendices_Theorem_R_t_R_s_zone}.
\end{IEEEproof}
\end{theorem}
\vspace{3ex}
By combining $R_t^\star$ and $R_e^\star$, the corresponding $R_s$ is obtained as
\begin{align}
R_s^\star = \frac{2}{\ln2}\mathbb{W}_0\left (\frac{{\eta_{_L}^{\frac{1}{2}}}2^{-R_e^\star+1}}{{\eta_{_N}^{\frac{1}{2}}}{\pi^2}{\lambda_u}{H}} \right ).
\label{R_s_zone}
\end{align}

To gain more insights, we further present the following corollary.
\begin{corollary}\label{coro_D}
It follows that as the guard zone radius $D$ increases, the optimal rate parameter $R_t^\star$ first decreases and then increases, while $R_s^\star$ increases monotonically. Furthermore, as $D \to \infty$, we have
\begin{align}
R_t^\star = R_s^\star = \frac{2}{\ln2}\mathbb{W}_0\left (\frac{2{\eta_{_L}^{\frac{1}{2}}}}{{\eta_{_N}^{\frac{1}{2}}}{\pi^2}{\lambda_u}{H}} \right ).
\label{R_t_R_s_D}
\end{align}
In this case, the corresponding secrecy transmission capacity becomes zero.
\begin{IEEEproof}
See Appendix \ref{appendices_coro_D}.
\end{IEEEproof}
\end{corollary}
\vspace{3ex}

Corollary \ref{coro_D} shows that as $D$ becomes significantly large, the secrecy transmission capacity becomes zero. This is intuitive, as the set of actual UAV transmitters becomes a null set in this case. As a result, it is expected that the optimal guard zone radius $D$ should be a finite value.

Finally, by combining $R_t^\star$ and $R_s^\star$, together with the 2D search over $H$ and $D$, the desirable system configurations are obtained for the case with secrecy guard zone. During the 2D search, we choose a sufficient large value of $D$ as the upper bound for search.

\section{Numerical Results}
In this section, we present numerical results to validate the theoretical analysis above. Unless otherwise stated, in the following we set the security constraint $\epsilon=0.01$, the UAVs' (or equivalently the legitimate receivers') density $\lambda_u = 10^{-3}$, the eavesdroppers' density $\lambda_e = 10^{-3}$, and the elevation angle threshold ${\theta_{c}} = {\pi/4}$.

\subsection{Secrecy Communication Performance under Given UAV Altitude $H$ and Guard Zone Radius $D$}
\begin{figure}[!h]
\centering
  \includegraphics[width=7.5cm]{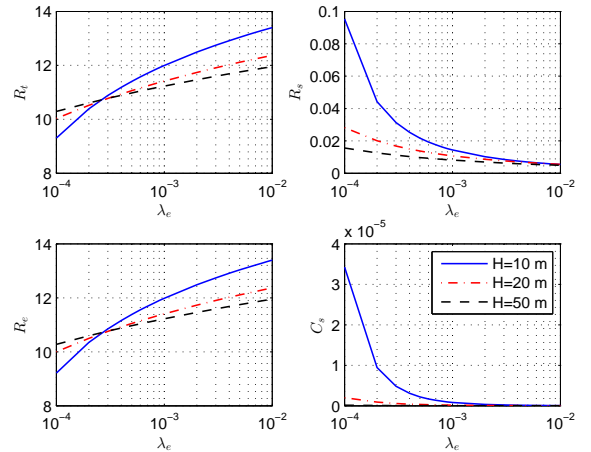}\\
  \caption{Secrecy communication performance without secrecy guard zone versus the eavesdroppers' density $\lambda_e$.}
\label{fig:fig1_fixH}
\end{figure}
First, we consider the case with given UAV altitude $H$ and guard zone radius $D$. Fig. \ref{fig:fig1_fixH} shows the secrecy communication performance without secrecy guard zone versus the eavesdroppers' density $\lambda_e$, where different values of UAV altitude $H$ are considered.
It is observed that as $\lambda_e$ increases, both $R_t$ and $R_e$ increase. This is because that in order to defend against more eavesdroppers, UAVs need to increase the randomness in the wiretap code (with increased $R_e$) to maintain the same level of secrecy. As a consequence, $R_t$ is increased as well.
Furthermore, as the attitude $H$ decreases, the values of $R_t$ and $R_e$ are observed to increase significantly. This is because that the eavesdropping channel quality becomes better when the attitude $H$ decreases, and thus $R_e$ should increase to add more randomness in the wiretap code to maintain the same level of secrecy.
It is also observed that as the UAV altitude $H$ decreases, the secrecy transmission capacity $\mathbb{C}_s$ increases.
This is intuitive, which is due to the fact that as $H$ decreases, the increase of the LoS legitimate channel power gains is more significantly than that of the eavesdropping channels.

\begin{figure}[!h]
\centering
  \includegraphics[width=7.5cm]{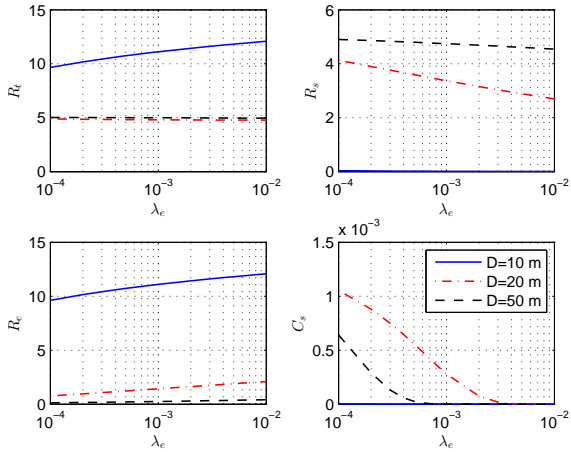}\\
  \caption{Secrecy communication performance with secrecy guard zone versus the eavesdroppers' density $\lambda_e$, where $H=20$~m.}
\label{fig:fig2_fixH_fixD}
\end{figure}
Fig. \ref{fig:fig2_fixH_fixD} shows the secrecy communication performance with secrecy guard zone versus the eavesdroppers' density $\lambda_e$, where we set $H=20$~m. Similar trends are observed as in Fig. \ref{fig:fig1_fixH}.
Besides, it is also observed that the guard zone radius $D = 20$~m leads to a larger secrecy transmission capacity $\mathbb{C}_s$ than $D = 10$ m and $D= 50$ m. This is consistent with our result in Corollary \ref{coro_D} that the optimal value of $D$ should be finite.

\begin{figure}[!h]
\centering
  \includegraphics[width=6.8cm]{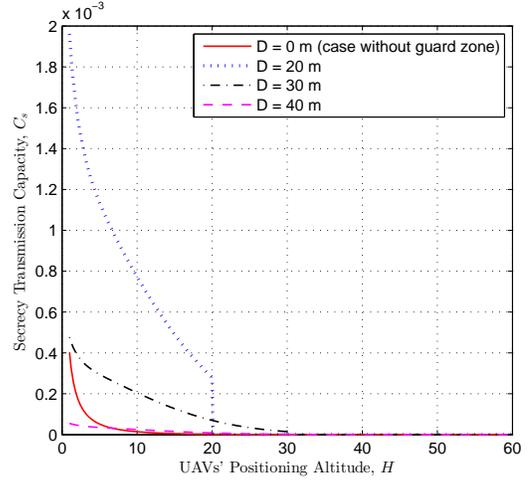}\\
  \caption{The secrecy transmission capacity $\mathbb{C}_s$ versus UAV positioning altitude $H$.}
\label{fig:optimal_H_fixD}
\end{figure}
Fig. \ref{fig:optimal_H_fixD} shows the secrecy transmission capacity $\mathbb{C}_s$ versus UAV positioning altitude $H$, under different values of guard zone radius $D$. It is observed that as the UAV altitude $H$ increases, the secrecy transmission capacity $\mathbb{C}_s$ decreases. This can be explained similarly as in Fig. \ref{fig:fig1_fixH}. When $D = 20$~m and $H \le 20$~m, the achieved secrecy transmission capacity is observed to significantly outperform that without guard zone employed (i.e., $D = 0$~m). This verifies the importance of the secrecy guard zone technique in secrecy communication performance enhancement. It is also observed that for all values of $D$, the secrecy transmission capacity decreases sharply as $H$ becomes larger than $D$.
This is expected, as in this case there exist eavesdroppers in the LoS regions (with LoS threshold $K = H$ under ${\theta_{c}} = {\pi/4}$), thus leading to stronger eavesdropping channels with reduced secrecy communication performance.

\subsection{Optimized Secrecy Transmission Design}
Next, we show the secrecy communication performance under optimized system configurations in terms of $R_t$, $R_s$, $H$, and $D$.

\begin{figure}[!h]
\centering
  \includegraphics[width=6.5cm]{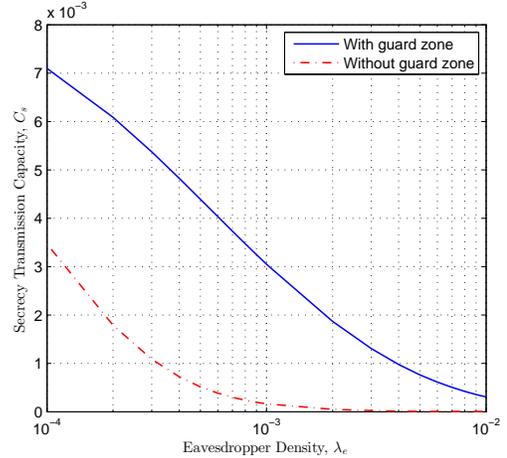}\\
  \caption{The optimized secrecy transmission capacity $\mathbb{C}_s$ versus the eavesdroppers' density $\lambda_e$.}
  \label{fig:optimal_C_lambda_e}
\end{figure}

\begin{figure}[!h]
\centering
  \includegraphics[width=6.5cm]{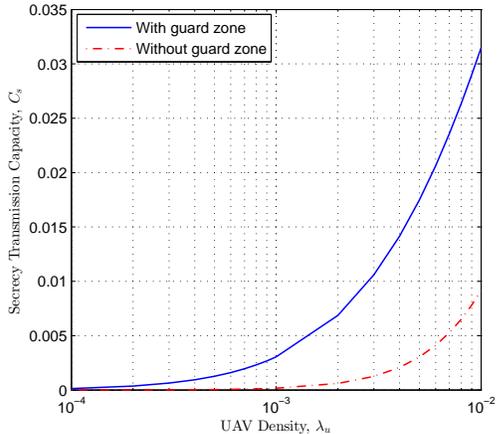}\\
  \caption{The optimized secrecy transmission capacity $\mathbb{C}_s$ versus the UAVs' density $\lambda_u$.}
  \label{fig:optimal_C_lambda_u}
\end{figure}
Figs. \ref{fig:optimal_C_lambda_e} and \ref{fig:optimal_C_lambda_u} show the optimized secrecy transmission capacity $\mathbb{C}_s$ versus the eavesdroppers' density $\lambda_e$ and the UAVs' density $\lambda_u$, respectively. It is observed that $\mathbb{C}_s$ decreases monotonically as $\lambda_e$ increases, while increases monotonically as $\lambda_u$ increases.
It is also observed that, under all values of $\lambda_e$, the use of secrecy guard zone leads to much higher secrecy transmission capacity than that without secrecy guard zone.

\begin{figure}[!h]
\centering
  \includegraphics[width=6.5cm]{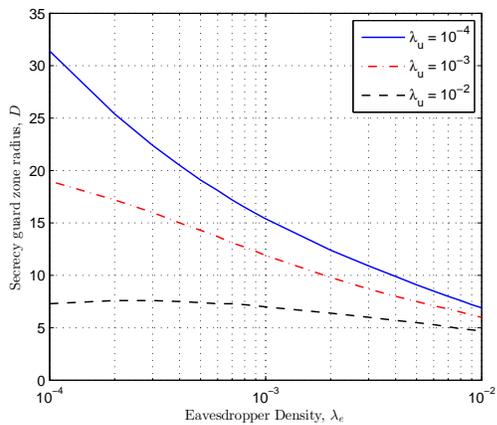}\\
  \caption{The optimized secrecy guard zone radius $D$ versus the eavesdroppers' density $\lambda_e$.}
  \label{fig:optimal_D_lambda_e}
\end{figure}
Fig. \ref{fig:optimal_D_lambda_e} shows the optimized secrecy guard zone radius $D$ versus the eavesdroppers' density $\lambda_e$. It is observed that the obtained $D$ decreases in general as $\lambda_e$ increases.
This shows that as the eavesdroppers' density increases, we should properly reduce $D$ to enable more UAVs to involve in the secrecy transmission, as the correspondingly achieved benefit outweighs the drawback caused by the increased chance of eavesdropping. It is also observed that the optimized guard zone radius $D$ decreases as $\lambda_u$ increases. This is expected, as the increase of UAVs' density leads to stronger interference at the eavesdroppers, thus leading to more significantly reduced SIR at eavesdroppers than legitimate receivers. In this case, a smaller guard zone radius $D$ is desirable.

\section{Concluding Remarks}
In this paper, we investigated the secrecy transmission of a large-scale UAV-enabled wireless network.
We model the horizontal locations of legitimate receivers and eavesdroppers as two independent PPPs and assume that the UAVs are each positioned exactly above the respective legitimate receivers for efficient secrecy communication. Under this setup and by considering two cases without and with the secrecy guard zone technique employed, we obtained analytical expressions for the connection probability and secrecy outage probability of this network in analytically tractable forms.
Accordingly, we optimized the system configurations (in terms of the Wyner's coding rates, UAV altitude, and guard zone size) to maximize the secrecy transmission capacity, subject to a maximum secrecy outage probability constraint.
Finally, we presented numerical results to validate the theoretical analysis, and show the effect of different system parameters on the secrecy communication performance.
Our results provided useful guidelines in analyzing the secrecy communication performance of large-scale UAV-enabled 3D wireless networks, and also provided insights on the design of secrecy transmission strategy and network configurations.
Due to the space limitation, there are still interesting problems unaddressed in this paper, which are briefly discussed in the following to motivate future work.
\begin{itemize}
\item This paper considered quasi-stationary UAV scenarios while in other mobile UAV scenario (i.e., the UAVs fly around during communication), how to analyze the network performance is a challenging task. One opinion to deal with this issue is to model the UAVs and the eavesdroppers as two independent random processes at each time instant. Then we can analyze the network performance at each time by using similarly techniques as in this paper, and then obtain the average performance by taking an expectation over time.
\item Furthermore, in mobile UAV scenarios, the UAVs may adjust their trajectories to optimize the secrecy communication performance. While prior work (see, e.g., \cite{ZhangSecuring2018,ZhongSecure2018}) considered the UAV trajectory optimization in the case with limited number of UAVs, how to combine such design with our large-scale network analysis for enhancing the performance a large-scale network is a challenging problem worth further investigation.
\item This paper considered the elevation-angle-dependent LoS/NLoS channels for A2G links based on the measurement results in \cite{KhawajaSurvey2019}, while there have been other A2G channel models in the literature such as the probabilistic LoS channels in \cite{HouraniOptimal2014} and elevation-angle-dependent Rician fading channels in \cite{AzariUltra2018}. How to analysis the large-scale network performance under different channel models is left for future work.
\item This paper considered that each UAV exactly hovers above one single ground user for efficient communication. For the scenario with UAVs acting as BSs to serve multiple users, each UAV may hover near these users but not exactly above them. In this case, the model in \cite{Liu2017Enhancing} may be applicable. How to extend our analysis in this case is interesting.
\item Additionally, there have been various practical issues and setups, such as different UAV attitude, directional antennas and full-duplex radio, which have not been considered in this work. How to extend our secrecy communication results in these scenarios is an interesting future direction.
\end{itemize}

\appendices
\section{Proof of Theorem \ref{Theorem_P_c}}\label{appendices_Theorem_P_c}
According to (\ref{P_c_1}), the connection probability can be computed as
\begin{align}
\label{P_c_2}{\mathcal{P}_{\textrm{c}}} &= {\mathcal{P}} \Bigg(\frac{{\eta_{_L}}{{{H}^{{-2} }}}}{\sum\limits_{u \in {\Phi_{u}}} {{{{\eta_{_{u0}}}{S_{u0}}}}{{{D_{u0}^{-\alpha_{u0}} }}}}} > \beta_t \Bigg)\\
\nonumber=& {\mathcal{P}} \Bigg(\frac{{{\eta_{_L}}}{{{H}^{{-2} }}}}{{\sum\limits_{l \in {\Phi_{L0}}} {{{{\eta_{_L}}}}{{{D_{l0}^{-{\alpha_{_L}}} }}}}}+\!\!\!\!{\sum\limits_{u \in {\Phi_{u}}\setminus{\Phi_{L0}}} {{{{\eta_{_N}}{{S}_N}}}{{{D_{u0}^{-{\alpha_{_N}}} }}}}}} > \beta_t \Bigg).
\end{align}

\newtheorem{lemma}{Lemma}
\begin{lemma}\label{lemma1}
According to \cite{Amari1997}, for a set of independent exponential random variables ${X} = \left\{ {{X_1}, \ldots ,{X_n}} \right\}$ with parameters ${\lambda _{{X_i}}},i = 1, \ldots ,n$, the cumulative distribution function (CDF) of the sum of independent exponentially distributed random variables $Y = \sum\limits_{i = 1}^n {{X_i}}$ is given by
\begin{align}
\mathcal{P}\left\{ {Y < y} \right\} = \sum\limits_{i = 1}^n {{\delta _i}\left( {1 - \exp \left[ { - {\lambda _{{X_i}}}y} \right]} \right)},
\label{SUM_EXP}
 \end{align}
where
\begin{align}
{\delta _i} = \prod\limits_{j = 1,j \ne i}^n {\frac{{{\lambda _{{X_j}}}}}{{{\lambda _{{X_j}}} - {\lambda _{{X_i}}}}}}.
\label{DELTA}
\end{align}
\end{lemma}
\vspace{3ex}

Applying Lemma \ref{lemma1}, (\ref{P_c_2}) can be re-expressed as (\ref{P_c_exact}). This theorem thus follows.

\section{Proof of Theorem \ref{Theorem_P_so_no_zone}}\label{appendices_Theorem_P_so_no_zone}
According to (\ref{P_so_1}), the secrecy outage probability can be computed as
\begin{align}
\label{P_so_1_free_space}{\mathcal{P}_{\textrm{so}}}=& {\mathcal{P}}\left ({\mathop {\max }\limits_{{e} \in {\Phi _{e}}} \left\{{{\tt SIR}_e} \right\}}> \beta_e \right)\\
\nonumber=&1-{\mathbb{E}_{\Phi_{u}}}\Bigg\{{\mathbb{E}_{\Phi _{e}}}\Bigg\{{\prod\limits_{{e} \in {\Phi _{e}} }}{\mathcal{P}}\left ({ {{\tt SIR}_e} }< \beta_e \right)\Bigg\}\Bigg\}.
\end{align}

According to \cite{Chiu2013}, the probability generating functional (PGFL) of a homogeneous PPP is given as
\begin{align}
{\mathbb{E}_{\Phi_e}}\left\{ {\prod\limits_{{e} \in {\Phi_e} } {f\left( {e} \right)} } \right\} =
 \exp \left[ { - \lambda_e \int_{{\mathbb{R}^2}} { {1 - f\left( e \right)}\, \text{d}e} } \right].
\label{PGFL}
\end{align}

Using PGFL of ${\Phi _{e}}$, (\ref{P_so_1_free_space}) can be computed as
\begin{align}
\nonumber{\mathcal{P}_{\textrm{so}}}&=1-{\mathbb{E}_{\Phi_{u}}}\Bigg\{\exp\Bigg[ { - {\lambda_{{e}}}} \int_{{\mathbb{R}^2}} {{\mathcal{P}}\left ({ {{\tt SIR}_e} }> \beta_e \right)}\, \text{d}e \Bigg]\Bigg\}\\
\label{P_so_2_free_space}&=1-{\mathbb{E}_{\Phi_{u}}}\Bigg\{\exp\Bigg[ { - {\lambda_{{e}}}}\\
\nonumber&\times\Bigg\{\underbrace{\int_{\boldsymbol{B}\left ( o, K \right )} {{\mathcal{P}}\left (\frac{{{\eta_{_L}}}{{D_{0e}^{-{\alpha_{_L}}} }}}{\sum\limits_{u \in {\Phi_{u}}} {{{{\eta_{_{ue}}}{S_{ue}}}}{{{D_{ue}^{-{\alpha_{ue}}} }}}}}> \beta_e \right)}\, \text{d}e}_{Q_2}\\
\nonumber+&\underbrace{\int_{{\mathbb{R}^2\setminus{\boldsymbol{B}\left ( o, K \right )}}} {{\mathcal{P}}\left (\frac{{{\eta_{_N}}{S_N}}{{D_{0e}^{-{\alpha_{_N}}} }}}{\sum\limits_{u \in {\Phi_{u}}} {{{{\eta_{_{ue}}}{S_{ue}}}}{{{D_{ue}^{-{\alpha_{ue}}} }}}}}> \beta_e \right)}\, \text{d}e}_{Q_3} \Bigg\}\Bigg]\Bigg\}.
\end{align}

Applying Lemma \ref{lemma1} over the typical eavesdropper $e \in {\Phi _{e}}$ lying inside ${\boldsymbol{B}\left ( o, K \right )}$, $Q_2$ in (\ref{P_so_2_free_space}) is computed as
\begin{align}\label{Q_2}
{Q_2} &= \int_{\boldsymbol{B}\left ( o, K \right )}\sum\limits_{u \in {\Phi_{u}}\setminus{\Phi_{Le}}}{\prod\limits_{{m \in {\Phi_{u}}\setminus{\Phi_{Le}}},m \ne u} {\frac{{{D_{me}^{\alpha_{_N}} }}}{{{{{D_{me}^{\alpha_{_N}} }}}  - {{{D_{ue}^{\alpha_{_N}} }}} }}} }\\
\nonumber&\times\left(1-\exp \left[ { - \frac{{\eta_{_L}}{D_{ue}^{\alpha_{_N}} }}{{\eta_{_N}}{\beta_e}{{D_{0e}^{\alpha_{_L}} }}}+\frac{\eta_{_L}{{D_{ue}^{\alpha_{_N}} }}}{\eta_{_N}}{\sum\limits_{l \in {\Phi_{Le}}} {D_{le}^{-\alpha_{_L} }}}   } \right]\right).
\end{align}

Also, by calculating two expectations over exponentially random variables ${S_N}$ and ${S_{ue}}$ when the typical eavesdropper $e \in {\Phi _{e}}$ lies outside ${\boldsymbol{B}\left ( o, K \right )}$, $Q_3$ in (\ref{P_so_2_free_space}) is computed as
\begin{align}\label{Q_3}
\nonumber{Q_3}=& \int_{{\mathbb{R}^2\setminus{\boldsymbol{B}\left ( o, K \right )}}} \!\!\!\!{{\mathbb{E}_{{S_{ue}}}}\left\{ {{\exp\left [-{\frac{{\beta_e}{D_{ue}^{\alpha_{_N}} }}{\eta_{_N}}\!\!{\sum\limits_{u \in {\Phi_{u}}} {{{{\eta_{_{ue}}}{S_{ue}}}}{{{D_{ue}^{-{\alpha_{ue}}} }}}}}} \right]} } \right\}}\\
\nonumber=&\int_{{\mathbb{R}^2\setminus{\boldsymbol{B}\left ( o, K \right )}}} {\prod\limits_{{l} \in {\Phi_{Le}} } {\left\{{\exp\left [-{\frac{{\eta_{_L}}{\beta_e}{{D_{0e}^{\alpha_{_N}} }}}{{\eta_{_N}} {{{{D_{le}^{\alpha_{_L}} }}}}}} \right]}\right\}} }\\
&\times{\prod\limits_{{u} \in {\Phi_{u}}\setminus{\Phi_{Le}} } {\left\{ \frac{1}{1+{{\beta_e}{{D_{0e}^{\alpha_{_N}} }}{{D_{ue}^{-{\alpha_{_N}}} }}}}\right\}} } \, \text{d}e \Bigg\}\Bigg]\Bigg\}.
\end{align}

By replacing ${Q_2}$ and ${Q_3}$ with (\ref{Q_2}) and (\ref{Q_3}), ${\mathcal{P}_{\textrm{so}}}$ in (\ref{P_so_2_free_space}) is re-expressed as (\ref{P_so_exact}). This theorem is thus verified.

\section{Proof of Corollary \ref{Corollary_P_c_approx}}\label{appendices_Corollary_P_c_approx}
According to (\ref{P_c_1_approx}), the approximate connection probability can be computed as
\begin{align}
\nonumber{\tilde{\mathcal{P}}_{\textrm{c}}} &= {\mathcal{P}}\left (\frac{{{\eta_{_L}}{\overline{S}_{00}}}{{{H}^{{-2} }}}}{\sum\limits_{u \in {\Phi_{u}}} {{{{\eta_{_{u0}}}{\overline{S}_{u0}}}}{{{D_{u0}^{-\alpha_{u0}} }}}}} > \beta_t \right)\\
\nonumber&={\mathbb{E}_{\Phi_{u}, {S_{u0}}}}\left\{ {{\exp\left [-{\frac{{\beta_t}{{{H}^{{2} }}}}{\eta_{_L}}{\sum\limits_{u \in {\Phi_{u}}} {{\eta_{_{u0}}}{{{\overline{S}_{u0}}}}{{D_{u0}^{-\alpha_{u0}}}}}}} \right]} } \right\}\\
&={\mathbb{E}_{\Phi_{u}}}\left\{ {\prod\limits_{{u} \in {\Phi_{u}} } {\left\{ \frac{1}{1+{{\beta_t}\eta_{_L}^{-1}{{{H}^{{2} }}}{\eta_{_{u0}}}{D_{u0}^{-\alpha_{u0}}}}}\right\}} } \right\}.
\label{P_c_2_approx}
\end{align}

Using PGFL of ${\Phi_{u}}$, (\ref{P_c_2_approx}) can be rewritten as
\begin{align}
{\tilde{\mathcal{P}}_{\textrm{c}}} = \exp\left[ { - {\lambda_{{u}}}\int_{{\mathbb{R}^2}} {1-\frac{1}{1+{{\beta_t}\eta_{_L}^{-1}{{{H}^{{2} }}}{\eta_{_{u0}}}{D_{u0}^{-\alpha_{u0}}}}} }\, \text{d}u} \right].
\label{P_c_3_approx}
\end{align}

Changing to polar coordinates, (\ref{P_c_3_approx}) can be turned to
\begin{align}
\label{P_c_4_approx}&{\tilde{\mathcal{P}}_{\textrm{c}}}= \exp\Bigg[ { - 2\pi\lambda_u }\\
\nonumber& \Bigg\{\int_{K}^{+\infty} {\left({1-\frac{1}{1+{{\beta_t}\eta_{_L}^{-1}{{{H}^{2 }}}{\eta_{_N}}\left({{r_u}^2+H^2}\right)^{-\frac{\alpha_{_N}}{2}}}} }\right) } {r_u}\, \text{d}r_u \\
\nonumber&+ \int_{0}^{K} {\left({1-\frac{1}{1+{{\beta_t}{{{H}^{2 }}}\left({{r_u}^2+H^2}\right)^{-\frac{\alpha_{_L}}{2}}}} }\right) } {r_u}\, \text{d}r_u \Bigg\}\Bigg].
\end{align}

Then, (\ref{P_c_4_approx}) can be re-expressed as (\ref{P_c_final}). This corollary thus follows.

\section{Proof of Corollary \ref{Corollary_P_so_no_zone_approx}}\label{appendices_Corollary_P_so_no_zone_approx}
According to (\ref{P_so_0_no_zone}), the approximate secrecy outage probability can be computed as
\begin{align}
\nonumber{\bar{\mathcal{P}}_{\textrm{so}}}=& {\mathcal{P}}\left ({\mathop {\max }\limits_{{e} \in {\Phi _{e}}} \left\{\frac{{{\eta_{_{0e}}}{\overline{S}_{0e}}}{{D_{0e}^{-\alpha_{0e}} }}}{\sum\limits_{u \in {\Phi_{u}}} {{{{\eta_{_{ue}}}{\overline{S}_{ue}}}}{{{D_{ue}^{-\alpha_{ue}} }}}}} \right\}}> \beta_e \right)\\
\label{P_so_1_no_zone}=&1-{\mathbb{E}_{\Phi_{u}}}\Bigg\{{\mathbb{E}_{\Phi _{e}}}\Bigg\{{\prod\limits_{{e} \in {\Phi_{e}} }}\\
\nonumber&\Bigg\{1- {\prod\limits_{{u} \in {\Phi_{u}} } {\left\{ \exp\left [-\frac{{{\eta_{_{ue}}}}{\beta_e}{\overline{S}_{ue}}{D_{ue}^{-{\alpha_{ue}} }}}{{{\eta_{_{0e}}}}{{D_{0e}^{-{\alpha_{0e}} }}}} \right]\right\}}} \Bigg\} \Bigg\}\Bigg\}.
\end{align}

According to (\ref{PGFL}), (\ref{P_so_1_no_zone}) can be rewritten as
\begin{align}
\label{P_so_2_no_zone}{\bar{\mathcal{P}}_{\textrm{so}}}=&1-{\mathbb{E}_{\Phi_{u}}}\Bigg\{\exp\Bigg[ { - {\lambda_{{e}}}}\\
\nonumber&\times \int_{{\mathbb{R}^2}} {\prod\limits_{{u} \in {\Phi_{u}} } {\left\{ \exp\left [-\frac{{{\eta_{_{ue}}}}{\beta_e}{\overline{S}_{ue}}{D_{ue}^{-{\alpha_{ue}} }}}{{{\eta_{_{0e}}}}{{D_{0e}^{-{\alpha_{0e}} }}}} \right]\right\}}}\, \text{d}e \Bigg]\Bigg\}.
\end{align}

Using Jensen's inequality, an upper bound of ${\bar{\mathcal{P}}_{\textrm{so}}}$ can be obtained:
\begin{align}
\label{P_so_3_no_zone}{\bar{\mathcal{P}}_{\textrm{so}}} \approx& 1-\exp\Bigg[ { - {\lambda_{{e}}}} \\
\nonumber\times& \int_{{\mathbb{R}^2}}{\mathbb{E}_{\Phi_{u}}}\left\{{\prod\limits_{{u} \in {\Phi_{u}} } {\left\{ \exp\left [-\frac{{{\eta_{_{ue}}}}{\beta_e}{\overline{S}_{ue}}{D_{ue}^{-{\alpha_{ue}} }}}{{{\eta_{_{0e}}}}{{D_{0e}^{-{\alpha_{0e}} }}}} \right]\right\}}}\right\}\, \text{d}e \Bigg].
\end{align}

Similar to (\ref{P_c_2_approx}), (\ref{P_so_3_no_zone}) can be computed as
\begin{align}
\label{P_so_4_no_zone}{\bar{\mathcal{P}}_{\textrm{so}}} &\approx 1-\exp\Bigg[ { - {\lambda_{{e}}}\int_{{\mathbb{R}^2}} \exp \Bigg[ { - {{\frac{{\pi}{\lambda_u}{D_{0e}^{\frac{\alpha_{0e}}{2}}}{\beta_e^{\frac{1}{2}}}{\eta_{_N}^{\frac{1}{2}}}}{2{\eta_{_{0e}}^{\frac{1}{2}}}}}}}}\\
\nonumber&\times \left(\pi-2\arctan\left({\frac{{\eta_{_{0e}}^{\frac{1}{2}}}\left(H^2+{K}^{2}\right)}{{\eta_{_N}^{\frac{1}{2}}}D_{0e}^{\frac{\alpha_{0e}}{2}}{\beta_e^{\frac{1}{2}}}}}\right) \right) \\
\nonumber&-{\pi}{\lambda_u}{{D_{0e}^{\alpha_{0e}}}{{\beta_e}}}\log\left(\frac{{{D_{0e}^{\alpha_{0e}}}{{\beta_e}}}+{H}^{2}+{K}^{2}}{{{D_{0e}^{\alpha_{0e}}}{{\beta_e}}}+{H}^{2}} \right)\Bigg]\, \text{d}e \Bigg].
\end{align}

In the case with $D_{0e} \ge K$, we have ${\frac{{\eta_{_{0e}}^{\frac{1}{2}}}\left(H^2+{K}^{2}\right)}{{\eta_{_N}^{\frac{1}{2}}}D_{0e}^{\frac{\alpha_{0e}}{2}}{\beta_e^{\frac{1}{2}}}}} \ll 1$ by combing the facts that ${\beta_e} \gg 1$, ${\eta_{_{0e}}} = \eta_{_N}$, and ${\alpha_{0e}} = 4$. In the other case with $D_{0e} < K$, the value of ${\frac{{\eta_{_{0e}}^{\frac{1}{2}}}\left(H^2+{K}^{2}\right)}{{\eta_{_N}^{\frac{1}{2}}}D_{0e}^{\frac{\alpha_{0e}}{2}}{\beta_e^{\frac{1}{2}}}}}$ is generally no larger than $1$, as ${\beta_e} \gg 1$, ${\eta_{_{0e}}} = \eta_{_L} \le \eta_{_N}$, and ${\alpha_{0e}} = 2$.
It should be noted that the probability that the typical eavesdropper $e$ lying inside the LoS region is much smaller than that lying inside the NLoS region.
Hence, it generally follows that ${\frac{{\eta_{_{0e}}^{\frac{1}{2}}}\left(H^2+{K}^{2}\right)}{{\eta_{_N}^{\frac{1}{2}}}D_{0e}^{\frac{\alpha_{0e}}{2}}{\beta_e^{\frac{1}{2}}}}} \ll 1$. Similarly, we have $\frac{{K}^{2}}{{{D_{0e}^{\alpha_{0e}}}{{\beta_e}}}+{H}^{2}} \ll 1$.
By adopting the approximation $\frac{{\eta_{_N}^{\frac{1}{2}}}D_{0e}^{\frac{\alpha_{0e}}{2}}{\beta_e^{\frac{1}{2}}}}{{{{\eta_{_{0e}}^{\frac{1}{2}}}\left(H^2+{K}^{2}\right)}}}\arctan\left({\frac{{\eta_{_{0e}}^{\frac{1}{2}}}\left(H^2+{K}^{2}\right)}{{\eta_{_N}^{\frac{1}{2}}}D_{0e}^{\frac{\alpha_{0e}}{2}}{\beta_e^{\frac{1}{2}}}}}\right)\approx 1$ and ${{D_{0e}^{\alpha_{0e}}}{{\beta_e}}}\log\left(\frac{{{D_{0e}^{\alpha_{0e}}}{{\beta_e}}}+{H}^{2}+{K}^{2}}{{{D_{0e}^{\alpha_{0e}}}{{\beta_e}}}+{H}^{2}} \right) \approx {{K}^{2}}$, we approximate (\ref{P_so_4_no_zone}) as
\begin{align}
\nonumber{\tilde{\mathcal{P}}_{\textrm{so}}} =& 1-\exp\Bigg[ { - {\lambda_{{e}}}}\int_{{\mathbb{R}^2}}\exp \Bigg[  - {{\frac{\pi}{2}{{\lambda_u}{D_{0e}^{\frac{\alpha_{0e}}{2}}}{\beta_e^{\frac{1}{2}}}}}} \\
&\times\left(\frac{{\pi}{\eta_{_N}^{\frac{1}{2}}}}{{\eta_{_{0e}}^{\frac{1}{2}}}}-{\frac{2 H^2}{D_{0e}^{\frac{\alpha_{0e}}{2}}{\beta_e^{\frac{1}{2}}}}} \right) \Bigg]\, \text{d}e \Bigg]
.\label{P_so_5_no_zone}
\end{align}

Changing to polar coordinates, (\ref{P_so_5_no_zone}) can be turned to
\begin{align}
\nonumber {\tilde{\mathcal{P}}_{\textrm{so}}}=& 1-\exp\Bigg[ { - 2\pi\lambda_e }\Bigg\{\int_{0}^{K}\exp \Bigg[  - {{\frac{{\pi}}{2}}}{\lambda_u}\left(r_e^{2}+{H^2}\right )^{\frac{\alpha_{_L}}{4}}\\
\nonumber&\times{\beta_e^{\frac{1}{2}}}\left(\frac{{\pi}{\eta_{_N}^{\frac{1}{2}}}}{{\eta_{_L}^{\frac{1}{2}}}}-{\frac{2 H^2}{\left(r_e^{2}+{H^2}\right )^{\frac{\alpha_{_L}}{4}}{\beta_e^{\frac{1}{2}}}}} \right) \Bigg]  {r_e}\, \text{d}r_e\\
\nonumber&+\int_{K}^{+\infty}\exp \Bigg[  - {{\frac{{\pi}}{2}}}{\lambda_u}\left(r_e^{2}+{H^2}\right )^{\frac{\alpha_{_N}}{4}}\\
&\times{\beta_e^{\frac{1}{2}}}\left(\pi-{\frac{2 H^2}{\left(r_e^{2}+{H^2}\right )^{\frac{\alpha_{_N}}{4}}{\beta_e^{\frac{1}{2}}}}} \right) \Bigg]  {r_e}\, \text{d}r_e\Bigg\}
\Bigg].
\label{P_so_6_no_zone}
\end{align}

Then, (\ref{P_so_6_no_zone}) can be simplified as (\ref{P_so_final_no_zone}). This corollary is thus verified.

\section{Proof of Theorem \ref{Theorem_R_t_R_s}}\label{appendices_Theorem_R_t_R_s}
The first derivative of $\mathbb{C}_s$ with respect to $R_t$ is computed as
\begin{align}
\nonumber \frac{\mathrm{d} \mathbb{C}_s}{\mathrm{d} {R_t}} =&{\left [ 1-{{{\frac{{\ln2}}{4}}}{\pi^2}{\lambda_u}{H}}\;{2^{\frac{R_t}{2}}}{\frac{{\eta_{_N}^{\frac{1}{2}}}}{{\eta_{_L}^{\frac{1}{2}}}}}\left ( R_t-R_e^* \right ) \right ]}\\
 &\times{\exp \left[ { - {{\frac{{\pi}}{2}}}{\lambda_u}{H}\left({\frac{{\eta_{_N}^{\frac{1}{2}}}}{{\eta_{_L}^{\frac{1}{2}}}}}\pi{2^{\frac{R_t}{2}}}-2{H} \right)} \right]}{\lambda_u}.
\label{R_t_first_derivative1_1}
\end{align}

Next, the second derivative of $\mathbb{C}_s$ with respect to $R_t$ is computed as
\begin{align}\label{R_t_first_derivative2_1}
&\frac{\mathrm{d}^2 \mathbb{C}_s}{\mathrm{d} {R_t}^2}= \frac{{\ln2}\;{{\pi^2}{\lambda_u}{H}}\;2^{\frac{R_t}{2}}{\eta_{_N}^{\frac{1}{2}}}}{4{\eta_{_L}^{\frac{1}{2}}}}\\
\nonumber&\;\;\times{\exp \left[ { - {{\frac{{\pi}}{2}}}{\lambda_u}{H}\left({\frac{{\eta_{_N}^{\frac{1}{2}}}}{{\eta_{_L}^{\frac{1}{2}}}}}\pi{2^{\frac{R_t}{2}}}-2{H} \right)} \right]}\\
\nonumber&\;\;\times\left [ -2-\frac{\ln2}{2}\left ( R_t-R_e^* \right )\left ({\frac{{\eta_{_N}^{\frac{1}{2}}}}{{\eta_{_L}^{\frac{1}{2}}}}} 2^{{\frac{R_t}{2}}-1}{{\pi^2}{\lambda_u}{H}}-1 \right ) \right ]{\lambda_u}.
\end{align}
It is easy to obtain that $\frac{\mathrm{d}^2 \mathbb{C}_s}{\mathrm{d} {R_t}^2} < 0$. Therefore, it follows that $\mathbb{C}_s$ is concave in $R_t \ge 0$. In this case, by setting the first derivative being zero, we have $R_t^* = {R_e^*} + \frac{2}{\ln2}\mathbb{W}_0\left (\frac{{\eta_{_L}^{\frac{1}{2}}}2^{-R_e^*+1}}{{\eta_{_N}^{\frac{1}{2}}}{\pi^2}{\lambda_u}{H}} \right )$.

Therefore, $R_t^*$ is optimal for maximizing $\mathbb{C}_s$. This theorem is thus proved.

\section{Proof of Theorem \ref{Theorem_P_so}}\label{appendices_Theorem_P_so}
According to (\ref{P_so_zone_1}), using PGFL of ${\Phi _{e}}$, the secrecy outage probability is computed as
\begin{align}
\nonumber{\mathcal{P}_{\textrm{so}}^{\textrm{zone}}}&=1-{\mathbb{E}_{\Phi_{u}}}\Bigg\{\exp\Bigg[ { - {\lambda_{{e}}}} \int_{{\mathbb{R}^2}\setminus{\boldsymbol{B}\left ( o, D \right )}} \!\!\!\!{{\mathcal{P}}\left ({ {{\tt SIR}_e} }> \beta_e \right)}\, \text{d}e \Bigg]\Bigg\}\\
\label{P_so_1_fading}&=1-{\mathbb{E}_{\Phi_{u}}}\Bigg\{\exp\Bigg[ { - {\lambda_{{e}}}} \\
\nonumber&\times\Bigg\{\int_{\boldsymbol{B}\left ( o, K \right )\setminus{\boldsymbol{B}\left ( o, D \right )}} \!\!\!\!\!\!\!\!{{\mathcal{P}}\left (\frac{{{\eta_{_L}}{P_t}}{{D_{0e}^{-{\alpha_{_L}}} }}}{\sum\limits_{u \in {\Phi_{u}}} {{{{\eta_{_{ue}}}{P_t}{S_{ue}}}}{{{D_{ue}^{-{\alpha_{ue}}} }}}}}> \beta_e \right)}\, \text{d}e\\
\nonumber&+\int_{{\mathbb{R}^2\setminus{\boldsymbol{B}\left ( o, K \right )}}} \!\!\!\!\!{{\mathcal{P}}\left (\frac{{{\eta_{_N}}{P_t}{{S}_N}}{{D_{0e}^{-{\alpha_{_N}}} }}}{\sum\limits_{u \in {\Phi_{u}}} {{{{\eta_{_{ue}}}{P_t}{S_{ue}}}}{{{D_{ue}^{-{\alpha_{ue}}} }}}}}> \beta_e \right)}\, \text{d}e \Bigg\}\Bigg]\Bigg\}.
\end{align}

Applying Lemma \ref{lemma1}, (\ref{P_so_1_fading}) can be re-expressed as (\ref{P_so_exact_zone}). This theorem is thus verified.

\section{Proof of Corollary \ref{Corollary_P_so_approx}}\label{appendices_Corollary_P_so_approx}
Using PGFL of ${\Phi _{e}}$, the secrecy outage probability is computed as
\begin{align}
\label{P_so_1_zone}&{\bar{\mathcal{P}}_{\textrm{so}}^{\textrm{zone}}}=1-{\mathbb{E}_{\Phi_{u}}}\Bigg\{\exp\Bigg[ { - {\lambda_{{e}}}}\\
\nonumber\times& \int_{{\mathbb{R}^2\setminus{\boldsymbol{B}\left ( o, D \right )}}} \!\!{\prod\limits_{{u} \in {\Phi_{u}} } {\left\{ \exp\left[-\frac{{{\eta_{_{ue}}}}{\beta_e}{\overline{S}_{ue}}{D_{ue}^{-{\alpha_{ue}} }}}{{{\eta_{_{0e}}}}{{D_{0e}^{-{\alpha_{0e}} }}}} \right]\right\}}}\, \text{d}e \Bigg]\Bigg\}.
\end{align}

Using Jensen's inequality, an upper bound of ${\tilde{\mathcal{P}}_{\textrm{so}}}$ is obtained as
\begin{align}
\label{P_so_2_zone}&{\bar{\mathcal{P}}_{\textrm{so}}^{\textrm{zone}}} \approx 1-\exp\Bigg[ { - {\lambda_{{e}}}} \\
\nonumber\times& \int_{{\mathbb{R}^2\setminus{\boldsymbol{B}\left ( o, D \right )}}}\!\!\!\!\!\!\!\!\!{\mathbb{E}_{\Phi_{u}}}\left\{{\prod\limits_{{u} \in {\Phi_{u}} } {\left\{ \exp\left [-\frac{{{\eta_{_{ue}}}}{\beta_e}{\overline{S}_{ue}}{D_{ue}^{-{\alpha_{ue}} }}}{{{\eta_{_{0e}}}}{{D_{0e}^{-{\alpha_{0e}} }}}} \right]\right\}}}\right\}\, \text{d}e \Bigg].
\end{align}

Similar to (\ref{P_c_2}), (\ref{P_so_2_zone}) is thus computed as
\begin{align}
\nonumber\label{P_so_3_zone}{\bar{\mathcal{P}}_{\textrm{so}}^{\textrm{zone}}} \approx& 1-\exp\Bigg[ { - {\lambda_{{e}}}\int_{{\mathbb{R}^2\setminus{\boldsymbol{B}\left ( o, D \right )}}} \exp \Bigg[ { - {{\frac{{\pi}{\lambda_u}{D_{0e}^{\frac{\alpha_{0e}}{2}}}{\beta_e^{\frac{1}{2}}}{\eta_{_N}^{\frac{1}{2}}}}{2{\eta_{_{0e}}^{\frac{1}{2}}}}}}}}\\
&\times \left(\pi-2\arctan\left({\frac{{\eta_{_{0e}}^{\frac{1}{2}}}\left(H^2+{K}^{2}\right)}{{\eta_{_N}^{\frac{1}{2}}}D_{0e}^{\frac{\alpha_{0e}}{2}}{\beta_e^{\frac{1}{2}}}}}\right) \right) \\
\nonumber&-{\pi}{\lambda_u}{{D_{0e}^{\alpha_{0e}}}{{\beta_e}}}\log\left(\frac{{{D_{0e}^{\alpha_{0e}}}{{\beta_e}}}+{H}^{2}+{K}^{2}}{{{D_{0e}^{\alpha_{0e}}}{{\beta_e}}}+{H}^{2}} \right)\Bigg]\, \text{d}e \Bigg].
\end{align}

Since ${\beta_e} \gg 1$ and ${\frac{H^2+{K}^{2}}{D_{0e}^{\frac{\alpha_{0e}}{2}}}} \ll 1$, we can derive a closed-form upper bound of (\ref{P_so_3_zone}) as
\begin{align}
\nonumber{\tilde{\mathcal{P}}_{\textrm{so}}^{\textrm{zone}}} =& 1-\exp\Bigg[ { - {\lambda_{{e}}}}\int_{{\mathbb{R}^2\setminus{\boldsymbol{B}\left ( o, D \right )}}}\exp \Bigg[  - {{\frac{\pi}{2}{{\lambda_u}{D_{0e}^{\frac{\alpha_{0e}}{2}}}{\beta_e^{\frac{1}{2}}}}}} \\
&\times\left(\frac{{\pi}{\eta_{_N}^{\frac{1}{2}}}}{{\eta_{_{0e}}^{\frac{1}{2}}}}-{\frac{2 H^2}{D_{0e}^{\frac{\alpha_{0e}}{2}}{\beta_e^{\frac{1}{2}}}}} \right) \Bigg]\, \text{d}e \Bigg]
.\label{P_so_4_zone}
\end{align}

Changing to polar coordinates, when ${D} \geq K$, (\ref{P_so_4_zone}) can be turned to
\begin{align}
\nonumber {\tilde{\mathcal{P}}_{\textrm{so}}^{\textrm{zone}}}=& 1-\exp\Bigg[ { - 2\pi\lambda_e }\int_{D}^{+\infty}\exp \Bigg[  - {{\frac{{\pi}}{2}}}{\lambda_u}\left(r_e^{2}+{H^2}\right )\\
&\times{\beta_e^{\frac{1}{2}}}\left(\pi-{\frac{2 H^2}{\left(r_e^{2}+{H^2}\right ){\beta_e^{\frac{1}{2}}}}} \right) \Bigg]  {r_e}\, \text{d}r_e\Bigg].
\label{P_so_5_zone}
\end{align}
Similarly, when ${D} < K$, we have
\begin{align}
\nonumber {\tilde{\mathcal{P}}_{\textrm{so}}^{\textrm{zone}}}=& 1-\exp\Bigg[ { - 2\pi\lambda_e }\Bigg\{\int_{D}^{K}\exp \Bigg[  - {{\frac{{\pi}}{2}}}{\lambda_u}\left(r_e^{2}+{H^2}\right )^{\frac{\alpha_{_L}}{4}}\\
\nonumber&\times{\beta_e^{\frac{1}{2}}}\left(\frac{{\pi}{\eta_{_N}^{\frac{1}{2}}}}{{\eta_{_L}^{\frac{1}{2}}}}-{\frac{2 H^2}{\left(r_e^{2}+{H^2}\right )^{\frac{\alpha_{_L}}{4}}{\beta_e^{\frac{1}{2}}}}} \right) \Bigg]  {r_e}\, \text{d}r_e\\
\nonumber&+\int_{K}^{+\infty}\exp \Bigg[  - {{\frac{{\pi}}{2}}}{\lambda_u}\left(r_e^{2}+{H^2}\right )^{\frac{\alpha_{_N}}{4}}\\
&\times{\beta_e^{\frac{1}{2}}}\left(\pi-{\frac{2 H^2}{\left(r_e^{2}+{H^2}\right )^{\frac{\alpha_{_N}}{4}}{\beta_e^{\frac{1}{2}}}}} \right) \Bigg]  {r_e}\, \text{d}r_e\Bigg\}
\Bigg].
\label{P_so_5_2_zone}
\end{align}

Then, (\ref{P_so_5_zone}) can be simplified as (\ref{P_so_final_zone}). This corollary is thus verified.

\section{Proof of Theorem \ref{Theorem_R_t_R_s_zone}}\label{appendices_Theorem_R_t_R_s_zone}
According to Theorem \ref{Theorem_R_t_R_s}, we can show that secrecy transmission capacity $\mathbb{C}_s$ is a concave function with respect to the rate $R_t$. Accordingly, $R_e^\star$ satisfies the equality (\ref{P_so_zone_R_e}). Then, the optimal values of $R_t^\star$ and $R_s^\star$ to maximize $\mathbb{C}_s$ are given as
\begin{align}
R_t^\star = R_e^\star + \frac{2}{\ln2}\mathbb{W}_0\left (\frac{{\eta_{_L}^{\frac{1}{2}}}2^{-R_e^\star+1}}{{\eta_{_N}^{\frac{1}{2}}}{\pi^2}{\lambda_u}{H}} \right ),
\end{align}
and
\begin{align}
R_s^\star = \frac{2}{\ln2}\mathbb{W}_0\left (\frac{{\eta_{_L}^{\frac{1}{2}}}2^{-R_e^\star+1}}{{\eta_{_N}^{\frac{1}{2}}}{\pi^2}{\lambda_u}{H}} \right ).
\end{align}
This completes the proof.

\section{Proof of Corollary \ref{coro_D}}\label{appendices_coro_D}
First, we show that $R_s^\star$ is an increasing function with respect to $D$.
From the explicit expression of ${\tilde{\mathcal{P}}_{{\textrm{so}}}^{\textrm{zone}}}$ (\ref{P_so_final_zone}), it is observed that ${\tilde{\mathcal{P}}_{{\textrm{so}}}^{\textrm{zone}}}$ is a decreasing function with respect to $D$ and $R_e$, respectively. Notice that at the optimality, we have ${\tilde{\mathcal{P}}_{{\textrm{so}}}^{\textrm{zone}}} = \epsilon$. Therefore, it is evident that $R_e^\star$ is also a decreasing function with respect to $D$. Furthermore, from the explicit expression of $R_s^\star$ in (\ref{R_s_zone}), we can observe that $R_s^\star$ is a decreasing function with respect to $R_e^\star$. Then we can conclude that $R_s^*$ is an increasing function with respect to $D$.

Next, we show that $R_t^\star$ first decreases and then increases as $D$ increases. The first derivative of $R_t^\star$ with respect to $R_e^\star$ is computed as
\begin{align}
\frac{\mathrm{d} R_t^\star}{\mathrm{d} {R_e^\star}} =\frac{1-\mathbb{W}_0\left (\frac{{\eta_{_L}^{\frac{1}{2}}}2^{-R_e^\star+1}}{{\eta_{_N}^{\frac{1}{2}}}{\pi^2}{\lambda_u}{H}} \right )}{1+\mathbb{W}_0\left (\frac{{\eta_{_L}^{\frac{1}{2}}}2^{-R_e^\star+1}}{{\eta_{_N}^{\frac{1}{2}}}{\pi^2}{\lambda_u}{H}} \right )}.
\label{}
\end{align}
It is observed that as $R_e^\star\ge 0$ increases, $\frac{\mathrm{d} R_t^\star}{\mathrm{d} {R_e^\star}}$ increases and the corresponding value of $\frac{\mathrm{d} R_t^\star}{\mathrm{d} {R_e^\star}}$ is first less than $0$ and then larger than $0$.
As $R_t^\star\ge 0$, $R_t^*$ first decreases and then increases as $R_e^\star$ increases. Hence, we can conclude that $R_t^\star$ first decreases and then increases as $D$ increases.

As $D\rightarrow +\infty$, it is easy to observe that $R_e^*\rightarrow 0$. In this case, by replacing $R_e^*$ with $0$ into (\ref{R_t_zone}) and (\ref{R_s_zone}), we can obtain
\begin{align}
R_t^* = R_s^* = \frac{2}{\ln2}\mathbb{W}_0\left (\frac{2{\eta_{_L}^{\frac{1}{2}}}}{{\eta_{_N}^{\frac{1}{2}}}{\pi^2}{\lambda_u}{H}} \right ).
\label{R_t_R_s_D_proof}
\end{align}

As $D\rightarrow +\infty$, we have ${\lambda_{u}}'\rightarrow 0$. Hence, it follows that $\mathbb{C}_s \rightarrow 0$ as $D\rightarrow +\infty$ from (\ref{P5_a}).

This completes the proof.

\footnotesize
\bibliographystyle{IEEEtran}
\bibliography{TCOM-ArXiv}

\end{document}